\documentclass[notitlepage,showpacs,preprintnumbers,amsmath,amssymb,superscriptaddress,twocolumn,aps,pra,longbibliography,showkeys]{revtex4-1}

\usepackage{graphicx}
\usepackage{dcolumn}
\usepackage{bm}
\usepackage{epstopdf}
\usepackage[colorlinks=true,linkcolor=blue,citecolor=red]{hyperref}

\usepackage[T1]{fontenc} 

\newcommand{\rozmiartrzy}{0.45\textwidth}

\newcommand{\NO}{NO}
\newcommand{\NOA}{NO$_{0}$}
\newcommand{\NOB}{NO$_{1}$}
\newcommand{\NOC}{NO$_{2}$}

\newcommand{\CBO}{CBO}
\newcommand{\CBOA}{CBO$_{1}$}
\newcommand{\CBOB}{CBO$_{2}$}

\newcommand{\STO}{STO}
\newcommand{\STOA}{STO$_{1}$}
\newcommand{\STOB}{STO$_{2}$}

\newcommand{\FNO}{FNO}
\newcommand{\FNOA}{FNO$_{1}$}
\newcommand{\FNOB}{FNO$_{2}$}
\newcommand{\FNOC}{FNO$_{3}$}

\newcommand{\FCBO}{FCBO}
\newcommand{\FSTO}{FSTO}

\begin{document}

\preprint{\emph{Submitted to:} Physical Review E}

\title{Diversity of charge orderings in correlated systems}%
\author{Konrad Jerzy Kapcia}%
\email[corresponding author; e-mail: ]{konrad.kapcia@ifpan.edu.pl}%
\affiliation{Institute of Physics, Polish Academy of Sciences, Aleja Lotnik\'ow 32/46, PL-02-668 Warsaw, Poland}%
\affiliation{Institute of Nuclear Physics, Polish Academy of Sciences, ul. E. Radzikowskiego 152, PL-31-342 Krak\'{o}w, Poland}%
\author{Jan Bara\'nski}%
\email[e-mail: ]{jan.baranski@ifpan.edu.pl}
\affiliation{Institute of Physics, Polish Academy of Sciences, Aleja Lotnik\'ow 32/46, PL-02-668 Warsaw, Poland}%
\author{Andrzej Ptok}%
\email[e-mail: ]{aptok@mmj.pl}
\affiliation{Institute of Nuclear Physics, Polish Academy of Sciences, ul. E. Radzikowskiego 152, PL-31-342 Krak\'{o}w, Poland}%
\affiliation{Institute of Physics, Maria Curie-Sk\l{}odowska University, Plac M. Sk\l{}odowskiej-Curie 1, PL-20-031 Lublin, Poland}%

\date{October 4, 2017}

\begin{abstract}
The phenomenon associated with inhomogeneous distribution of electron density is known as a charge ordering.
In this work, we study the zero-bandwidth limit of the extended Hubbard model, which can be considered as a simple effective model  of charge ordered insulators.
It  consists of the on-site interaction $U$ and the intersite density-density interactions $W_1$ and $W_2$ between  nearest-neighbors and next-nearest neighbors, respectively.
We derived the exact ground state diagrams for different lattice dimensionalities and discuss effects of small finite temperatures in the limit of high dimensions.
In particular, we estimated the critical interactions for which new ordered phases emerge (laminar or stripe and four-sublattice-type).
Our analysis show that the ground state of the model is highly degenerated.
One of the most intriguing finding is that the nonzero temperature removes these degenerations.
\end{abstract}

\pacs{\\
	71.10.Fd --- Lattice fermion models (Hubbard model, etc.),\\ 
	71.45.Lr --- Charge-density-wave systems,\\
	64.75.Gh --- 	Phase separation and segregation in model systems (hard spheres, Lennard-Jones, etc.),\\ 
	71.10.Hf --- Non-Fermi-liquid ground states, electron phase diagrams and phase transitions in model systems}
\keywords{\\
	extended Hubbard model, atomic limit, charge-order, stripes, long-range interactions, phase diagrams}
\maketitle


\section{Formulation of the problem}

The solutions of the extended Hubbard model predicts an existence of the states with the inhomogeneous spatial distribution of electrons \cite{WignerPR1934,RobaszkiewiczPSSB1973,MicnasRMP1990,PietigPRL1999,AichhornPRB2004,AmaricciPRB2010,HuangPRB2014,Giovannetti2015,KapciaPRB2017}.
This phenomenon is known as the charge ordering and can be observed in variety of compounds, e.g., cuprates \cite{CominScience2015,NetoScience2015,PelcNatCom2016,CaiNatPhys2016}, multiferroics \cite{HsuNatCom2016,ParkPRL2017} and other intensively studied materials  \cite{YoshimiPRL2012,FrandsenNatCom2014,NovelloPRL2017}.
The simplest example of such an order is an alternate modulation of electron concentration on the biparticle lattice.
In such a setup one can distinguish two equivalent sublattices, where every site in each sublattice is occupied by the same number of particles.
This is so-called two sublattice assumption.
In more general systems in which longer-range interactions play an important role the two-sublattice solutions does not capture the full basis of charge ordered phases \cite{BakPRB1980,BakPRL1982,BakRPP1982,LeePRL2001,RademakerPRE2013}.
In order to properly describe these orderings one needs to take into account more than two inequivalent sites. 
The most conspicuous classes of such materials are those where laminar or stripe orderings appear, e.g., manganites \cite{SalamonRMP2001,SunPNAS2011}, cobaltites \cite{AndradePRL2012}, and other transition-metal compounds \cite{TranquadaPRL1994}.

The exact solution of the extended Hubbard model is still unknown, even in one dimension.
A good testing field for approximate solutions would be results obtained for simplified, but exactly solvable  model (in arbitrary dimensions).
The Hubbard model (without intersite interactions) has been exactly solved in one dimension \cite{LiebPRL1968,LiebPRL1989} and in the limit of infinite dimension \cite{Georges1996RMP,N1,N2}.
Another approach is to neglect the hopping term instead of longer-range interactions.
Similar conditions can be met in narrow band materials \cite{MicnasRMP1990}.
One of the main goals of the present work is to provide the exact solutions that can be compared with approximate results for more complex systems (such as models taking into account quantum fluctuations introduced by the hopping term).

In this work we investigate the extended Hubbard model in the atomic limit taking into account the next-nearest-neighbour density-density interactions.
We assume that the mean-field solutions (with an exact treatment of the on-site terms) are contained within the four-sublattice system.
This assumption is justified as long as we do not take into account interactions with  longer range than the next-nearest neighbour.
Phase diagrams in this approach are obtained for full range of model parameters.
We note that the presented ground state solutions are exact for arbitrary dimensionality of the lattice.
In a case of the high-dimension limit  ($d\rightarrow+\infty$) we also present the effects associated with finite temperature.
In addition, we discuss the qualitative differences of ordering range depending on dimensionality of the considered system.

The extended Hubbard model in the zero-bandwidth limit with interactions restricted to the second neighbors can be expressed as:
\begin{eqnarray}
\label{eq:hamUW}
\label{row:1} \hat{H} & = & U\sum_i{n_{i\uparrow}n_{i\downarrow}} + \frac{W_{1}}{2z_1}\sum_{\langle i,j\rangle_1}{n_{i}n_{j}} \\
& & +\frac{W_{2}}{2z_2}\sum_{\langle i,j\rangle_2}{n_{i}n_{j}} - \mu\sum_{i}{n_{i}}, \nonumber
\end{eqnarray}
where $c^{\dag}_{i\sigma}$ denotes the creation operator of an electron with spin $\sigma$ at the site $i$, $n_{i}=\sum_{\sigma}{n_{i\sigma}}$, $n_{i\sigma}=c^{\dag}_{i\sigma}c_{i\sigma}$,
$U$ is the on-site density interaction,
and $W_{1}$ and $W_{2}$ are the intersite density-density interactions between nearest neighbours (NNs)
and next-nearest neighbours (NNNs), respectively.
$z_1$ and $z_2$ are numbers of NNs and NNNs, respectively.
$\mu$ is the chemical potential determining the total concentration
$n$ of electrons in the system by the relation
$n = \frac{1}{L}\sum_{i}{\left\langle n_{i} \right\rangle}$,
where \mbox{$0\leq n \leq 2$} and $L$ is the total number of lattice sites.
We  inspect phase diagrams emerging from this model.
The analyses are performed in the grand canonical ensemble.

Model (\ref{eq:hamUW}) with neglected next-nearest interactions ($W_2=0$) was intensively studied using various methods.
In particular, the exact solutions for a one-dimensional ($d=1$) chain were obtained using the transfer-matrix method \cite{BariPRB1971,BeniPRB1974,TuPSSB1974} or equations of motion and Green's function formalism \cite{ManciniEPL2005,ManciniEPJB2005,ManciniPRE2008,Mancini2009b}.
The rigorous ground state phase diagrams as a function of $\mu$ were obtained  for $W_2=0$ and $d\geq1 $ \cite{JedrzejewskiPhysA1994,BorgsJPA1996,FrohlichCMP2001}.
For a square lattice  ($d=2$) Monte Carlo simulations were performed \cite{PawlowskiEPJB2006,Ganzenmuller2008}.
The model on the Bethe lattice was also analysed~\cite{Mancini2009a,ManciniEPJB2010}.
The only known work beyond mean-field approaches for $W_2\neq 0$ treats the model within the transfer-matrix method for $d=1$ chain \cite{RobaszkiewiczPhysBC1982,ManciniEPJB2013} and the checker-board estimate with respect to lattice planes \cite{JedrzejewskiPhysA1994}.

Using the mean-field method for alternate lattices it was shown that the system can exhibit several homogeneous charge-ordered phases as well as various phase separated states \cite{MicnasPRB1984,BursillJPA1993,KapciaJPCM2011,KapciaPhysA2016}.
A case of $W_2\neq0$ within a mean-field approach was investigated in Refs.~\cite{KapciaJPCM2011,KapciaPhysA2016}.
However, these analyses were restricted to the two-sublattice assumption.
This restriction is sufficient only for attractive $W_2$, where there are no physical mechanisms supporting the four-sublattice type order.
In the present work we perform the studies of the model for the full range of parameters including repulsive $W_2>0$.
Our preliminary results only for $U<0$ have been presented in Ref.~\cite{KapciaJSNM2016}.
One of the conclusions of the mentioned work was that the magnitude of on-site attractive $U$ does not change the diagrams of the model qualitatively.
Therefore, in the present work we mainly focus on repulsive $U$.
We show that at particular values of $U>0$ new phases emerge.


\subsection*{The mean-field expressions}

The grand thermodynamic potential per site $\Omega_0$ for model (\ref{eq:hamUW}) in the grand canonical ensemble and in the mean-field four-sublattice approximation at $T=0$ can be expressed as
\begin{equation}
\label{eq:grandpotential.tempzero}
\Omega_0 = \langle \hat{H} \rangle / L  = E_{U} + E_W + E_\mu,
\end{equation}
where
\begin{eqnarray}
\label{eq:ED.tempzero}
E_{U} & = & \tfrac{1}{8} U  \left[ n_A (n_A-1) + n_B (n_B-1) \right. \\
&+ & \left. n_C (n_C-1) + n_D (n_D-1)  \right], \nonumber \\
\label{eq:EW.tempzero}
E_W & =& \frac{1}{8}W_1(n_A n_B +n_A n_D +n_B n_C + n_C n_D)\\
&+&\frac{1}{4}W_2(n_A n_C+n_B n_D), \nonumber \\
\label{eq:Emu.tempzero}
E_\mu & = & - \tfrac{1}{4} \mu  (n_A + n_B + n_C + n_D),
\end{eqnarray}
and $n_\alpha$ denotes the average number of particles in each sublattice $n_\alpha = \tfrac{4}{L} \sum_{i\in\alpha} \langle {n}_i \rangle$, $\alpha \in \{A,B,C,D\}$.
In this work we adopted the convention that the NNNs for any site from $A$ ($B$) sublattice  are those and only those sites, which are located in $C$ ($D$, respectively) sublattice.

For finite temperatures, the expressions given in Ref.~\cite{KapciaJPCM2011} in the four-sublattice assumption takes the following forms.
For $n_\alpha$ one gets
\begin{equation}
\label{eq:nalpha.fintemp}
n_\alpha = \tfrac{2}{Z_\alpha} \left[ \exp{ \left( \beta  \mu_\alpha \right) } + \exp{\left( 2\mu_\alpha - U \right)} \right].
\end{equation}
For a grand canonical potential (per lattice site) one obtains:
\begin{equation}
\label{eq:grandpotential.fintemp}
\Omega = - \frac{1}{8}\sum_{\alpha}\Phi_\alpha n_\alpha -\frac{1}{4\beta} \sum_{\alpha} \left( \ln{Z_\alpha} \right),
\end{equation}
where $\beta=1/(k_BT)$ is inverted temperature, $\mu_\alpha = \mu - \Phi_\alpha$, and
\begin{eqnarray}
\label{eq:Zalpha.fintemp}
Z_\alpha& =& 1 + 2 \exp \beta \mu_\alpha + \exp{\beta \left( 2\mu_\alpha - U \right)}, \\
\label{eq:phiA.fintemp}
\Phi_A & = & \tfrac{1}{2} W_1 (n_B+ n_D) + W_2 n_C,\\
\label{eq:phiB.fintemp}
\Phi_B & = & \tfrac{1}{2} W_1 (n_A+ n_C) + W_2 n_D,\\
\label{eq:phiC.fintemp}
\Phi_C & = & \tfrac{1}{2} W_1 (n_B+ n_D) + W_2 n_A,\\
\label{eq:phiD.fintemp}
\Phi_D & = & \tfrac{1}{2} W_1 (n_A+ n_C) + W_2 n_B.
\end{eqnarray}

\begin{figure}
	\includegraphics[width=0.49\textwidth]{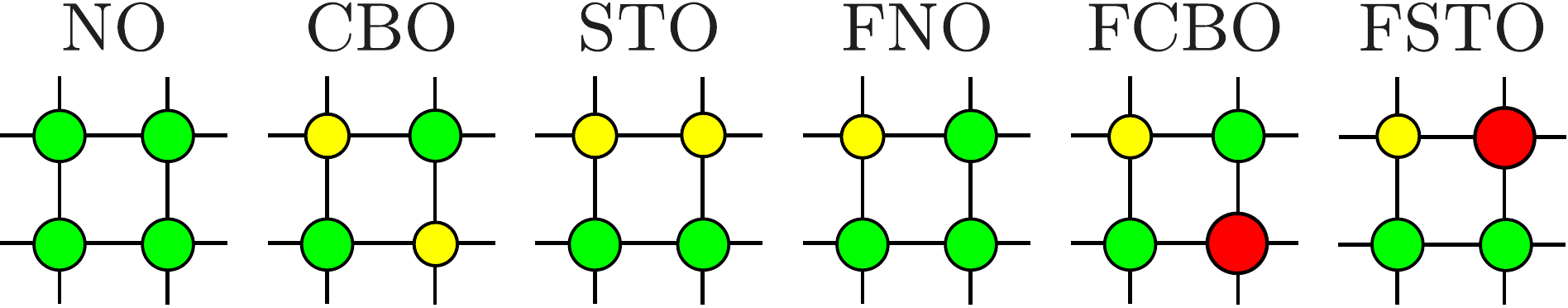}
	\caption{
		Schematic representation of all possible homogeneous solutions for four-sublattice orderings at the ground state.
		Different sizes of dots correspond to different concentrations $n_\alpha$ ($\alpha=A,B,C,D$) in the sublattices.
		Note that  each pattern can be realised in a few distinct forms (Tables~\ref{tab:funkmi} and \ref{tab:concentartion}) due to specific concentrations of each site.
	}
	\label{fig:schem}
\end{figure}

\section{Phase diagrams at high dimensions}

In this paragraph we present the ground state solutions of model (\ref{eq:hamUW}) in the limit of high dimensions i.e. $d\rightarrow+\infty$ (or equivalently the limit of large coordination number: $z_1\rightarrow+\infty$ and $z_2\rightarrow+\infty$)  and compare them to the results obtained for finite temperatures.
We present the phase diagrams for fixed  chemical potential as well as for fixed total electron concentration.
In the following analyses $|W_1|$ is used as the energy unit.
The solutions for repulsive $W_1>0$ and attractive $W_1<0$ are analysed separately.

In our system only six (inequivalent) homogeneous phases can occur at $T=0$.
They are determined by the relations between $n_\alpha$'s  (but several equivalent solutions exist due to cyclic change of sublattices indexes $\alpha$).
For intuitive understanding of rather complicated phase diagrams each pattern is marked with adequate abbreviation.
Nonordered (\NO), checker-board-ordered (\CBO), and stripe-ordered (\STO) phases can be realized using two sublattices, while the letter ``F'' (in the names of the {\FNO}, {\FCBO}, and {\FSTO} phases) indicates that these types of ordering requires the four-sublattice assumption.
All these phases are schematically depicted in Fig.~\ref{fig:schem}.
Each pattern can be realised in a few distinct forms depending on specific electron concentrations on each sublattice (cf. Tables~\ref{tab:funkmi} and \ref{tab:concentartion}).
Table~\ref{tab:funkmi} also contains the degeneracy of the ground state solutions (including charge- and spin- degrees of freedom).

\subsection{Analysis for fixed chemical potential for repulsive $W_1$}
\label{sec:MFAW1positivechempot}

First, we focus on the case of repulsive $W_1>0$.
We discuss the qualitative changes of a phase diagrams with respect to on-site Coulomb interaction ($U$) as a function of chemical potential and next-nearest neighbour intersite interactions ($W_2$).
Below we present phase diagrams for a few representative on-site interactions, where qualitative differences occur.
The diagrams are plotted for a full range of NNN interaction $W_2$ and shifted chemical potential $\bar{\mu}$  ($\bar{\mu} = \mu- \tfrac{1}{2}U-W_1-W_2$).
Notice that the model exhibits the particle-hole symmetry and thus all phase diagrams are symmetric towards $\bar{\mu} = 0$ (or $n=1$) with $n_\alpha \longleftrightarrow 2-n_\alpha$ transformation  if one changes $\bar{\mu}\longleftrightarrow-\bar{\mu}$ (or $n\longleftrightarrow 2-n$).
The qualitative changes of phase diagrams occur at $U=0.00$, $U=0.50|W_1|$, and $U=1.00|W_1|$.
All possible phases (within the four-sublattice assumption) are collected in Table~\ref{tab:funkmi}.
Their grand canonical potentials $\Omega_0$ are calculated from Eqs. (\ref{eq:grandpotential.tempzero})--(\ref{eq:Emu.tempzero}) and the phase with the lowest $\Omega_0$ for given model parameters is determined.

It is rather intuitive that the attractive on-site interactions favours double occupancy and only phases with empty and doubly occupied states
(i.e., the {\NOA} (0000), {\CBOB} (2020), {\STOB} (2200), and {\FNOB} (2000) phases) are stable for $U<0$ (the numbers in the brackets denote the concentrations in sublattices ($n_A n_B n_C n_D$)).
Changing the sign of $U$ leads to the appearance of new phases with single occupied sites.
These states can be considered as intermediate phases emerging from boundaries,
e.g., from boundary between {\NOA} (0000) and {\FNOB} (2000) phases the intermediate {\FNOA} (1000) phase emerges.
A Similar case occurs on the other side of the {\FNOB} phase region, where for $U>0$ the {\FSTO} (2100) phase appears as an intermediate one between the {\FNOB} and {\STOB} (2200) phases.
In such conditions the {\FNOB} phase is sandwiched between two intermediate growing phases.

The qualitative changeover occurs for on-site energy $U = 0.50|W_1|$.
For this value of interaction two electrons occupying one site in the {\FNOB} phase can be separated to form the stripe-ordered {\STOA} (1100) phase and thus for larger $U$ the {\STOA} phase is stable.
We note that additional two phases (the {\FNOC} (1110) and {\NOB} (1111) phases) without double occupied sites emerge on boundaries around $W_2=0.5|W_1|$ and $-0.25|W_1|<\bar{\mu}<0$.
For $U>0.5|W_1|$ the {\FNOC} and {\NOB} phases occurs in finite range of model parameters.
With further increasing of $U$ the region of the {\FCBO}  (2010) phase shrinks.
The region of this phase occurrence at $U=1.00|W_1|$ is reduced to a single point.
For $U>1.00|W_1|$ the {\NOB}--{\CBOA} transitions appears and for larger values of $U$ the diagram does not change,  qualitatively.
Remarkably, the {\FCBO}  phase exists only for define range of on-site interaction, while the {\FSTO} phase is present for arbitrarily large $U$.

\begin{figure*}[t!]
        \centering
        \includegraphics[width=\rozmiartrzy]{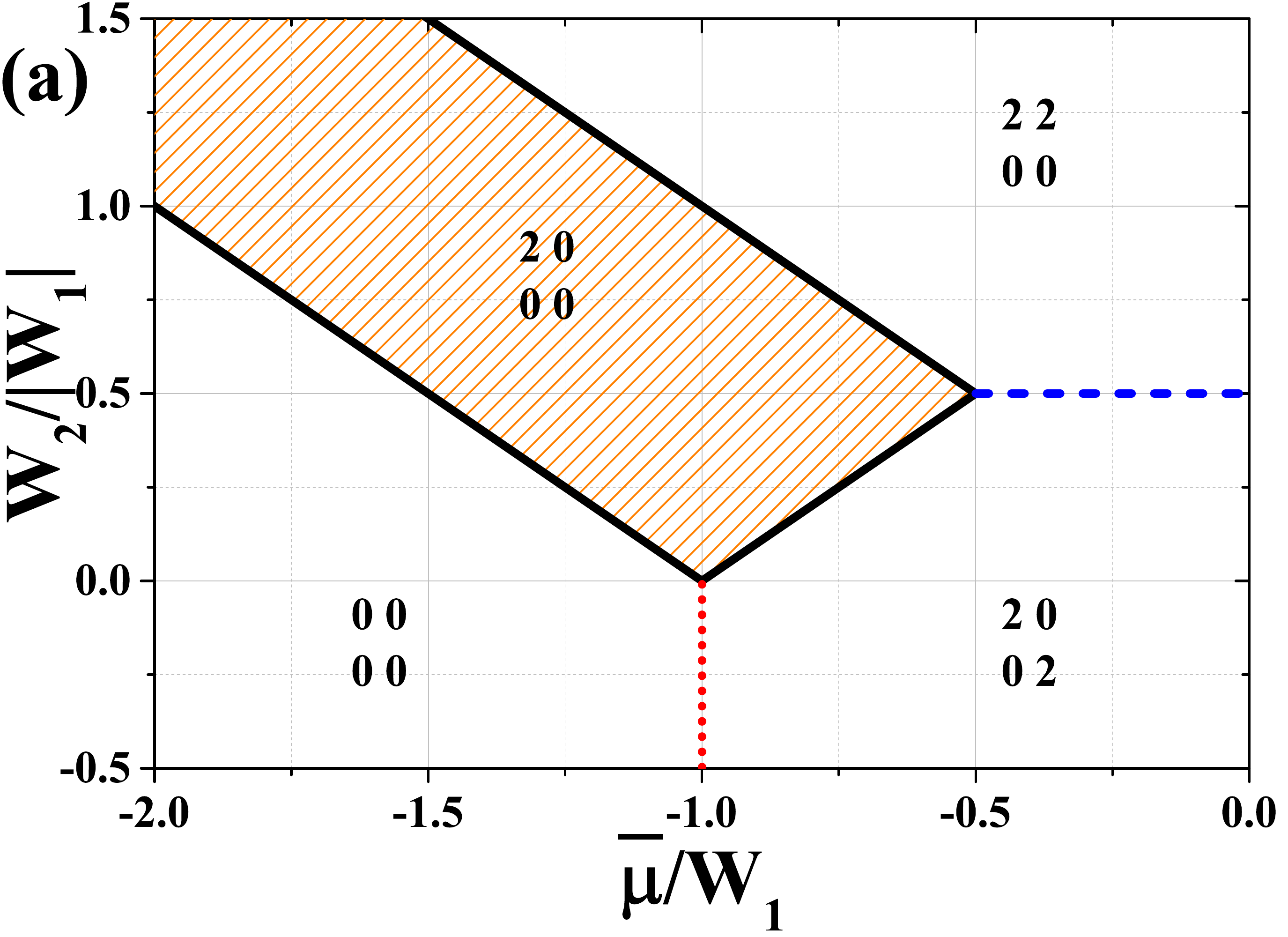}
        \includegraphics[width=\rozmiartrzy]{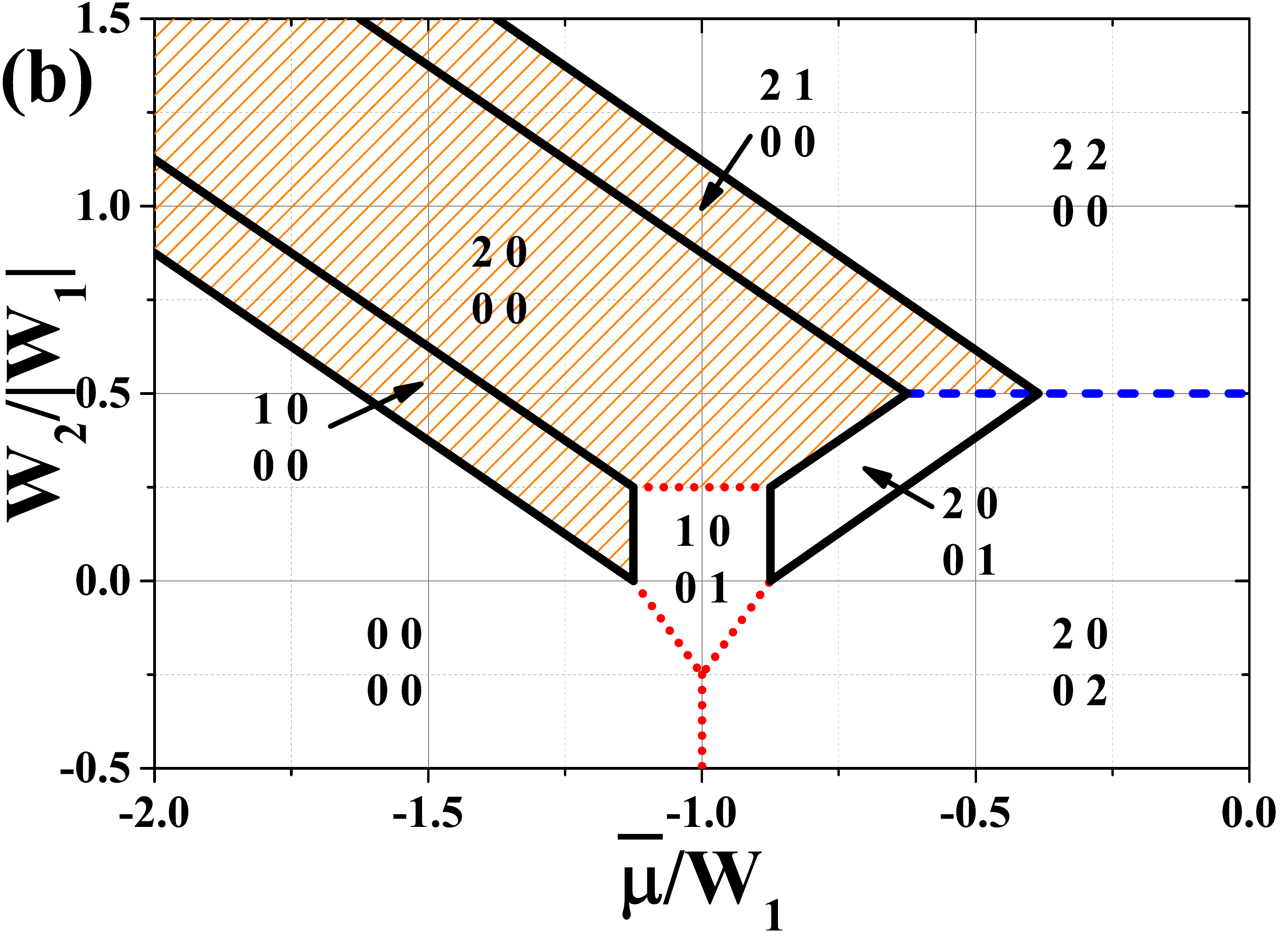}\\
        \includegraphics[width=\rozmiartrzy]{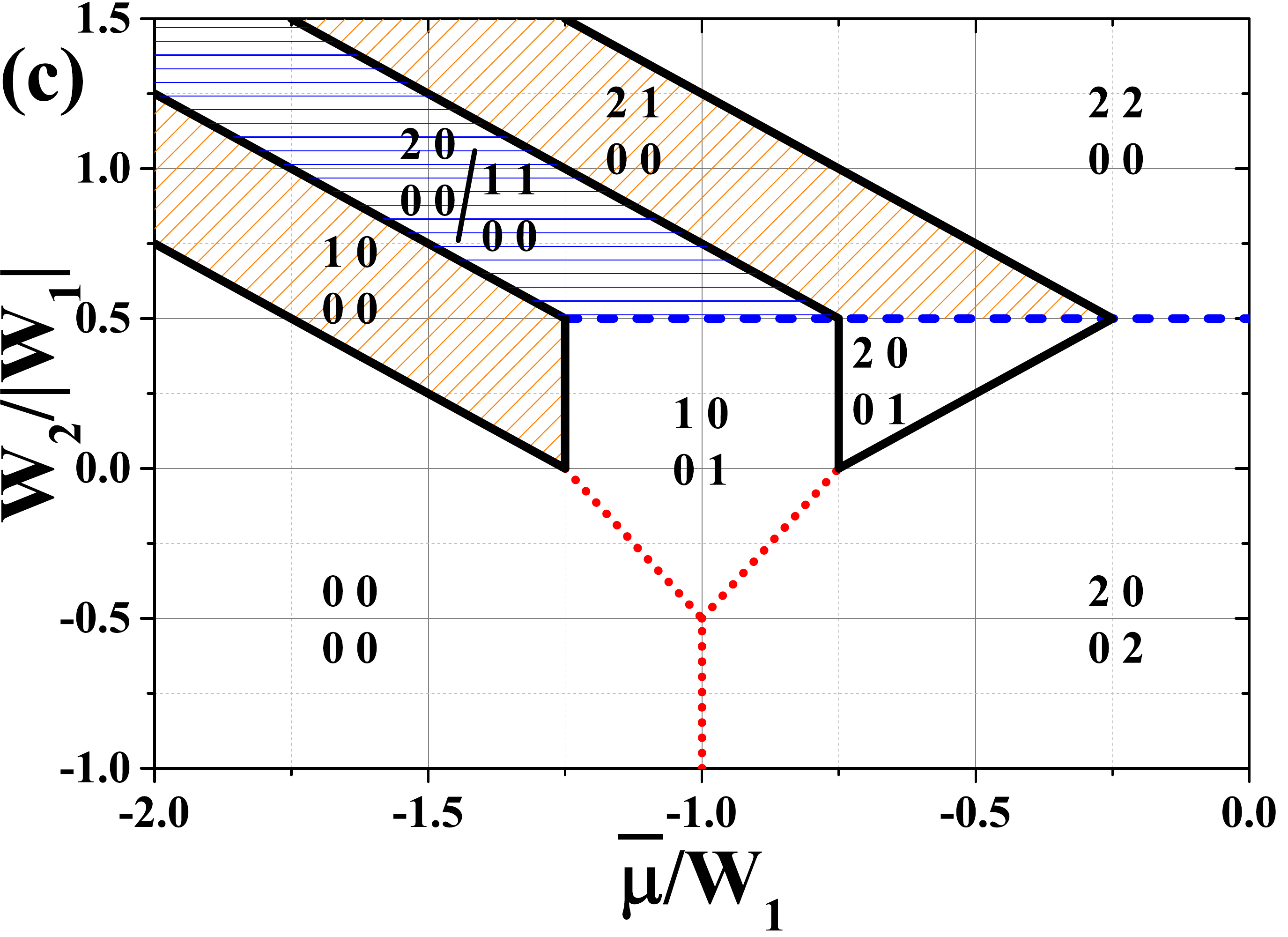}
        \includegraphics[width=\rozmiartrzy]{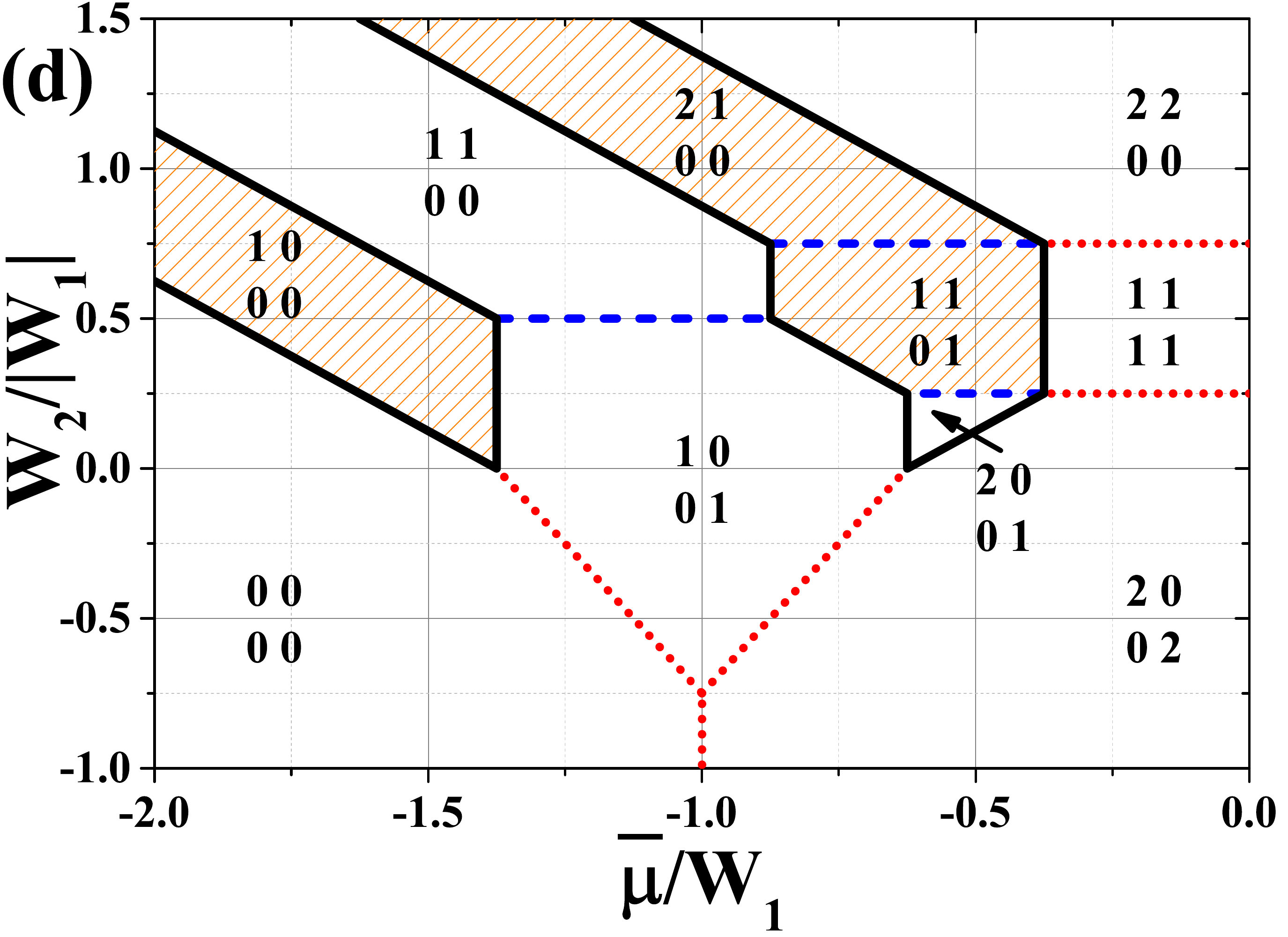}\\
        \includegraphics[width=\rozmiartrzy]{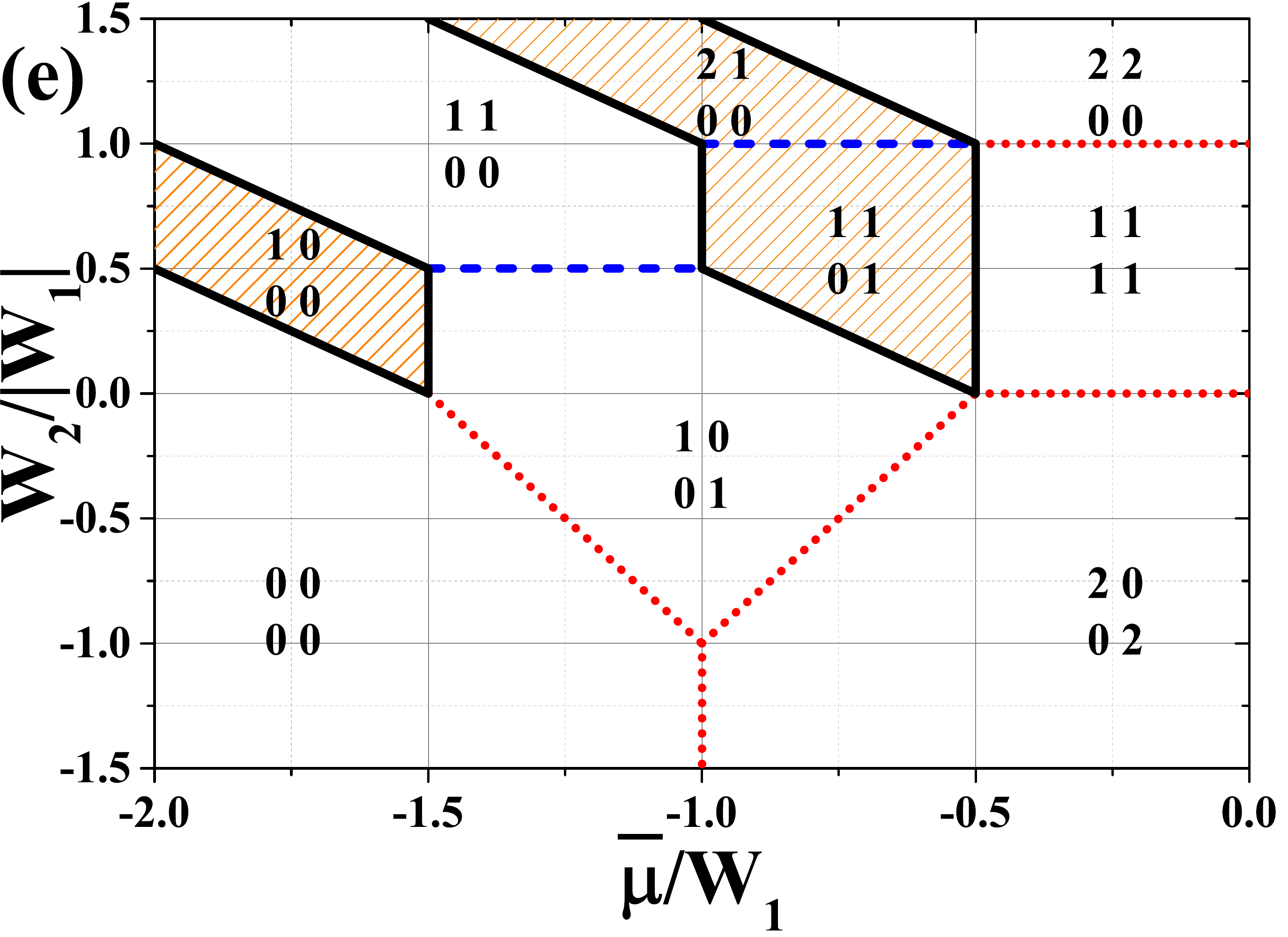}
        \includegraphics[width=\rozmiartrzy]{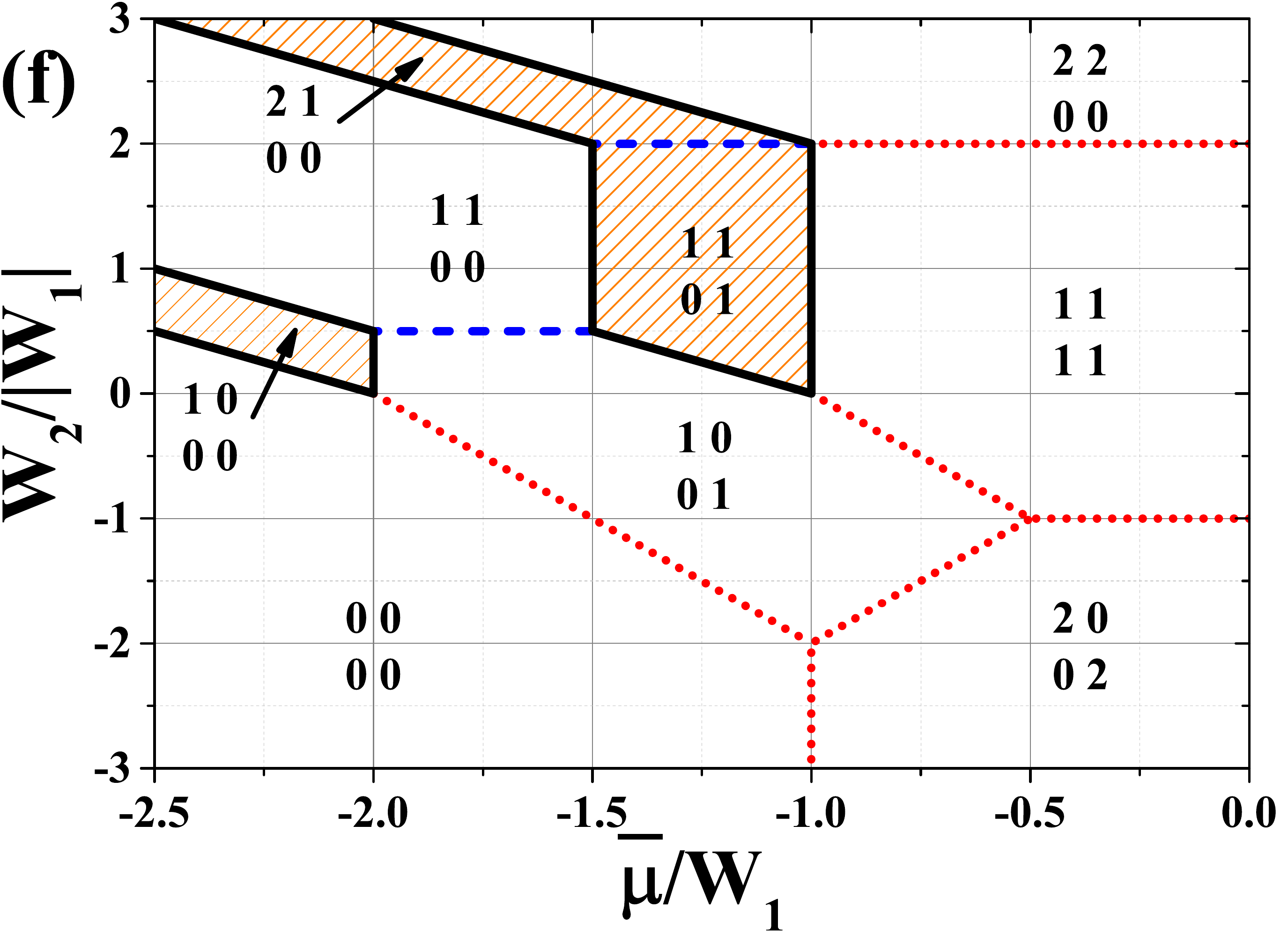}
        \caption{(Color online)
        Ground state phase diagrams as a function of chemical potential for $W_1>0$ and (a) $U/|W_1|\leq0.00$, (b) $U/|W_1|=0.25$, (c) $U/|W_1|=0.50$, (d) $U/|W_1|=0.75$, (e) $U/|W_1|=1.00$, and (f) $U/|W_1|=2.00$ ($\bar{\mu} = \mu- \tfrac{1}{2}U-W_1-W_2$).
        Each region is labelled by electron concentrations in each sublattice, 	$\left( \begin{smallmatrix} n_A & n_B \\ n_D & n_C \end{smallmatrix} \right)$ (cf. Table~\ref{tab:funkmi}).
        Regions filled by a slantwise pattern: ``reduced long-range order'' in $d=1,2$.
        At the boundaries for $d=2$: solid lines -- macroscopic degeneration, dashed lines (and inside the region filled by horizontal pattern) -- infinite degeneration but not macroscopic, dotted lines -- finite degeneration (modulo spin).
        }
        \label{rys:GSchemical}
\end{figure*}

\begin{table}
	\caption{%
		The definitions of phases ($d\leq3$) or elementary blocks ($d=1,2$) occurring in the ground state for total $n\leq1$.
		Notation: $2$ -- doubly occupied site, $0$ -- empty site, $1$ -- singly occupied site.
		The degeneration $d_c \times d_s$ of the elementary blocks (equal to the degeneration of the ground state for $d\rightarrow+\infty$ limit) and degeneration $D_c \times D_s$ of the ground state phases constructed from the corresponding blocks for $d=2$ ($L=N^2$) is given (with respect to charge- and spin- degrees of freedom).
		}
		\label{tab:funkmi}
	\begin{ruledtabular}
		\begin{tabular}{l|cccc|lc}
			Phase & $n_A$ & $n_B$ & $n_C$ & $n_D$ &  $d_c \times d_s$ & $D_c \times D_s$ \\
			\hline
		\NOA	& $0$ & $0$ & $0$ & $0$ & $1  \times 1$  &  $1  \times 1$  \\
		\NOB		& $1$ & $1$ & $1$ & $1$ & $1  \times 16$  &  $ 1   \times 2^{L}$  \\
		\hline
		\CBOA	& $1$ & $0$ & $1$ & $0$ & $2  \times 4$  &  $2  \times 2^{L/2}$  \\
		\CBOB	& $2$ & $0$ & $2$ & $0$ & $2  \times 1$  &  $2  \times 1$  \\
		\hline
		\STOA		& $1$ & $1$ & $0$ & $0$ & $4  \times 4$  &  $4  \times 2^{L/2}$  \\
		\STOB	& $2$ & $2$ & $0$ & $0$ & $4  \times 1$  &  $4  \times 1$  \\
		\hline
		\FNOA \footnotemark[1]		& $1$ & $0$ & $0$ & $0$ & $4  \times 2$  &  $2^{N/2+2}   \times 2^{L/4}$  \\
		\FNOB \footnotemark[1]	& $2$ & $0$ & $0$ & $0$ & $4  \times 1$  &  $2^{N/2+2}  \times 1$  \\
		\FNOC \footnotemark[1]		& $1$ & $1$ & $1$ & $0$ & $4  \times 8$  &  $ 2^{N/2+2}   \times 2^{3L/4}$  \\
		\hline
		\FCBO 		& $2$ & $0$ & $1$ & $0$ & $4  \times 2$  &  $ 4    \times 2^{L/4}$  \\
		\FCBO'\footnotemark[2]		& $2$ & $1$ & $0$ & $1$ & $4  \times 4$  &  $ 4 \times 2^{L/2}$  \\
		\hline
		\FSTO \footnotemark[1]		& $2$ & $1$ & $0$ & $0$ & $8  \times 2$  &  $ 2^{N/2+2}    \times 2^{L/4}$  \\
		\FSTO'\footnotemark[1]${^{,}}$\footnotemark[2]		& $2$ & $1$ & $1$ & $0$ & $4  \times 4$  &  $2^{N/2+2}  \times 2^{L/2}$  \\
		\end{tabular}
		\footnotetext[1]{The long-range charge-order is ``reduced'' in the phase constructed from this elementary blocks in $d=1,2$.}
		\footnotetext[2]{This phase constructed from this elementary block does not occur in the ground state.}
	\end{ruledtabular}
\end{table}

Notice that all boundaries between ground state phases in Fig.~\ref{rys:GSchemical} are discontinuous (first order transitions associated with discontinuous change of at least one of the $n_\alpha$'s).
At the boundaries two phases possess the same energy.
If two phases coexist in the system the interfaces between them are formed.
To determine the state of the system in such conditions one needs to estimate the additional energy required for formation of the phase interface.
If creation of the interface between two phases does not require additional energy, then these two phases can coexist on a microscopic level.
In other words the four-site building blocks of each phases can be aligned arbitrary next to each other.
More detailed discussion of this issue is included in Section \ref{sec:lowdim}.

The situation is more complex for nonzero interface energies. 
In this case the phases cannot be mixed on a microscopic level.
Nevertheless, if the size of the interface increases as $L^{\gamma}$ with $\gamma < 1$ (where $L$ is the number of lattice sites) the contribution of the interface energy to the total energy of the system vanishes in the thermodynamic limit.
In such a case, the macroscopic phase domains will be formed.
A formation of these regions with different orderings is known in physics as the (macroscopic) phase separation (PS) (the only one possible type of a coexistence of two phases in $d\rightarrow+\infty$ limit).
Otherwise,
for $\gamma>1$ the formation of the interfaces is disfavoured and the co-existence of the phases is not allowed in the system.
In such situation, even though both phases have the same energy, only one type of the ordering is realized in all volume of the system.
We note, however, that we did not find such behaviour in the considered model.

\begin{table*}
	\caption{%
		The definitions of homogeneous phases  which can occur in the mean-field ground state of the model as a function of $n$  and ranges $\left[n_s, n_f\right]$ of electron concentration where the phases are defined properly.
		For each phase the double occupancy $D_{occ}$ defined as
		$D_{occ}=\tfrac{1}{L}\sum_i\left\langle {n}_{i\uparrow} {n}_{i\downarrow} \right\rangle $
		is also calculated (exact result for $d\rightarrow+\infty$).
		At the last column the corresponding phase separated states are mentioned (cf. Table~\ref{tab:funkmi}).}
	\label{tab:concentartion}
	\begin{ruledtabular}
		\begin{tabular}{lcccccccl}
			Phase & $n_A$ & $n_B$ & $n_C$ & $n_D$ & $D_{occ}$ & $n_s$ & $n_f$ & PS state \\
			\hline
			\NO$_{\textrm{A}}$ & $n$  & $n$ & $n$ & $n$ & $n/2$ & $0$ & $2$ & \NOA/\NOC \\
			\NO$_{\textrm{B}}$ & $n$  & $n$ & $n$ & $n$ & $0$ & $0$ & $1$ & \NOA/\NOB \\
			\hline
			\CBO$_{\textrm{A}}$  & $2n$ & $0$ & $2n$ & $0$ & $n/2$     & $0$ & $1$ & \NOA/\CBOB \\
			\CBO$_{\textrm{B}}$  & $2n$ & $0$ & $2n$ & $0$ & $0$       & $0$ & $1/2$ & \NOA/\CBOA \\
			\CBO$_{\textrm{C}}$  & $2n$ & $0$ & $2n$ & $0$ & $n-1/2$   & $1/2$ & $1$ & \CBOA/\CBOB \\
			\CBO$_{\textrm{D}}$  & $1$ & $2n-1$ & $1$ & $2n-1$ & $0$ & $1/2$ & $1$ & \CBOA/\NOB \\
			\hline
			\STO$_{\textrm{A}}$ & $2n$ & $2n$ & $0$ & $0$ & $n/2$ & $0$ & $1$ & \NOA/\STOB \\
			\STO$_{\textrm{B}}$ & $2n$ & $2n$ & $0$ & $0$ & $0$ & $0$ & $1/2$ & \NOA/\STOA \\
			\STO$_{\textrm{C}}$ & $2n$ & $2n$ & $0$ & $0$ & $n-1/2$ & $1/2$ & $1$ & \STOA/\STOB \\
			\STO$_{\textrm{D}}$ & $1$ & $2n-1$ & $2n-1$ & $1$ & $0$ & $1/2$ & $1$  & \STOA/\NOB \\
			\hline
			\FNO$_{\textrm{A}}$  & $4n$ & $0$ & $0$ & $0$ &  $n/2$     & $0$ & $1/2$ & \NOA/\FNOB \\
			\FNO$_{\textrm{B}}$  & $4n$ & $0$ & $0$ & $0$ &  $0$       & $0$ & $1/4$ & \NOA/\FNOA \\
			\FNO$_{\textrm{C}}$  & $4n$ & $0$ & $0$ & $0$ &  $n-1/4$   & $1/4$ & $1/2$ & \FNOA/\FNOB \\
			\hline
			\FCBO$_{\textrm{A}}$  & $2$ & $0$ & $4n-2$ & $0$ & $n/2$ & $1/2$ & $1$ & \FNOB/\CBOB\\
			\FCBO$_{\textrm{B}}$  & $2$ & $0$ & $4n-2$ & $0$ & $n-1/2$ & $3/4$ & $1$ & \FCBO/\CBOB\\
			\FCBO$_{\textrm{C}}$  & $2$ & $0$ & $4n-2$ & $0$ & $1/4$ & $1/2$ & $3/4$ & \FNOB/\FCBO\\
			\FCBO$_{\textrm{D}}$  & $1$     & $0$   & $4n-1$    & $0$   & $0$       & $1/4$ & $1/2$ & \FNOA/\CBOA \\
			\FCBO$_{\textrm{E}}$  & $1$     & $0$   & $4n-1$    & $0$   & $n-1/2$   & $1/2$ & $3/4$ & \CBOA/\FCBO\\
			\FCBO$_{\textrm{F}}$  & $1$ & $4n-2$ & $1$ & $0$ & $0$ & $1/2$ & $3/4$ & \CBOA/\FNOC\\
			\FCBO$_{\textrm{G}}$   & $1$ & $ 4n-3$ & $1$ & $1$ & $0$ & $3/4$ & $1$ & \FNOC/\NOB\\
			\hline
			\FSTO$_{\textrm{A}}$  & $2$ & $4n-2$  & $0$ & $0$ & $n/2$ & $1/2$ & $1$ & \FNOB/\STOB\\
			\FSTO$_{\textrm{B}}$  & $2$ & $4n-2$  & $0$ & $0$ & $1/4$ & $1/2$ & $3/4$ &  \FNOB/\FSTO\\
			\FSTO$_{\textrm{C}}$  & $2$ & $4n-2$  & $0$ & $0$ & $n-1/2$ & $3/4$ & $1$ & \FSTO/\STOB\\
			\FSTO$_{\textrm{D}}$  & $1$     & $4n-1$    &   0   & $0$   & $0$       & $1/4$ & $1/2$ & \FNOA/\STOA \\
			\FSTO$_{\textrm{E}}$  & $1$     & $4n-1$    &   0   & $0$   & $n-1/2$   & $1/2$ & $3/4$ & \STOA/\FSTO\\
			\FSTO$_{\textrm{F}}$  & $1$ & $1$ & $4n-2$ & $0$ & $0$ & $1/2$ & $3/4$ & \STOA/\FNOC\\
		\end{tabular}
	\end{ruledtabular}
\end{table*}

\subsubsection*{Effects of finite temperatures (fixed $\mu$)}

So far we considered the ground state ($T=0$), where all phase transitions are associated with a discontinuous change of at least one sublattice concentration $n_{\alpha}$.
In this subsection we discuss the influence of a finite temperatures on phase transitions slightly above the ground state.
The phases are found by numerically solving the set of four self-consistent equations in a form of (\ref{eq:nalpha.fintemp}) and finding the solution corresponding to the lowest $\Omega$ determined by (\ref{eq:grandpotential.fintemp}).

For attractive $W_2$ the behaviour of the system does not change qualitatively at small temperatures,
and all phase transitions remain first-order.
At critical $W_2=0$ the order of all transitions changes into the second one.
A more complex situation occurs for repulsive $W_2$.
Our analysis shows that all ground state {\FNO}--{\STO} boundaries (namely, (2000)--(2200), (1000)--(1100), and (1110)--(1100)) and boundary lines which are not dependent on the chemical potential (horizontal lines in Fig.~\ref{rys:GSchemical}) remain first-order.
All other transitions for $W_2>0$ are second-order at small $T>0$.
At finite temperature, the {\FNO} phases are no longer stable, and they are converted into {\FCBO} phases, i.e. $(n_A000)_{T=0}\rightarrow(n_An_Bn_Cn_B)_{T>0}$.
As a result the ground state {\FNOA}--{\FNOB} ((1000)--(2000)) and {\FNOB}--{\FCBO} ((2000)--(2010)) boundaries no longer exist at $T>0$ (Fig.~\ref{rys:GSchemical}(b)).
The similar situation takes place for the {\CBOA--\CBOB} boundary for $W_2=0$ \cite{KapciaPhysA2016}.

Additionally, we noticed that finite temperature, for small $W_2>0$, gives rise to checker-board orderings between {\NO} and {\FNO} phases.
Namely,  {\NO--\FNO} boundaries  change into the {\NO--\CBO--\FNO} sequence of second-order transitions with changing chemical potential.
In particular, the following sequences emerge: {\NOA--\CBOB--\FNOB},  {\NOA--\CBOA--\FNOA},  {\FNOC--\CBOA--\NOB} (cf. Ref. \cite{KapciaJSNM2016} for the $U<0$ case).
For larger values of $W_2$, the {\CBO} phases are absent and only  second-order transitions between corresponding {\NO} and {\FNO} phases occur.

\begin{figure*}[t!]
	\centering
	\includegraphics[width=\rozmiartrzy]{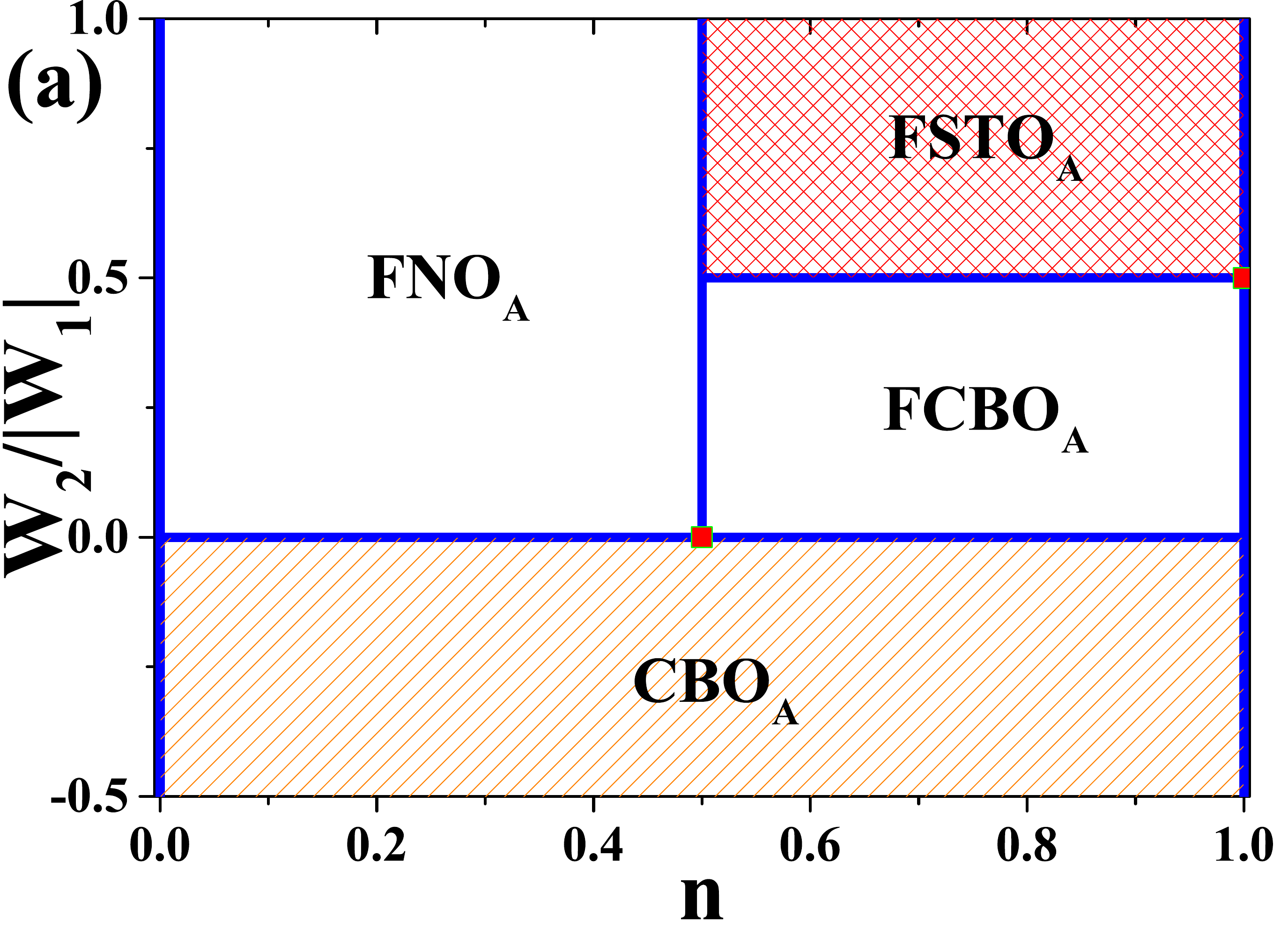}
	\includegraphics[width=\rozmiartrzy]{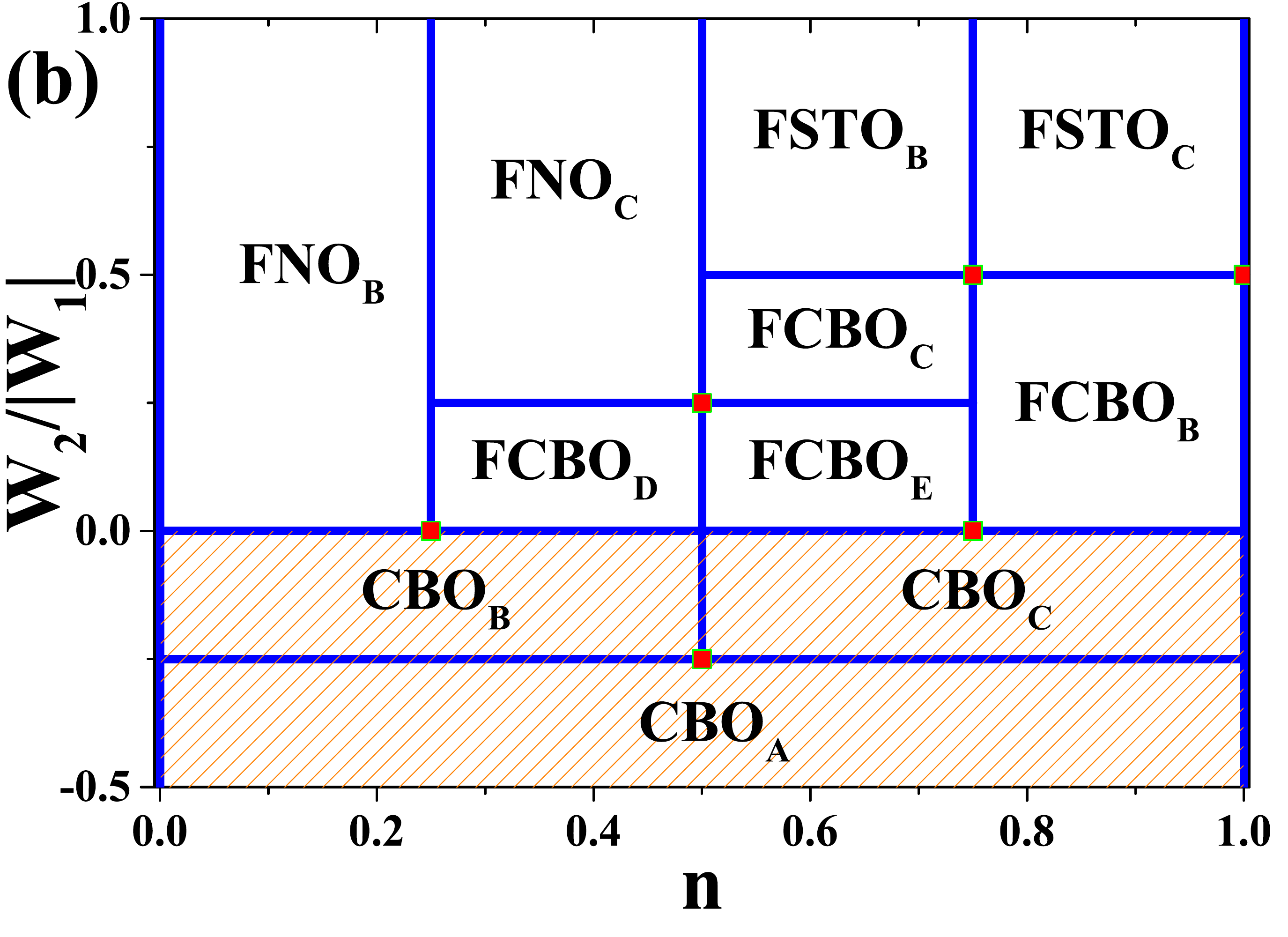}\\
	\includegraphics[width=\rozmiartrzy]{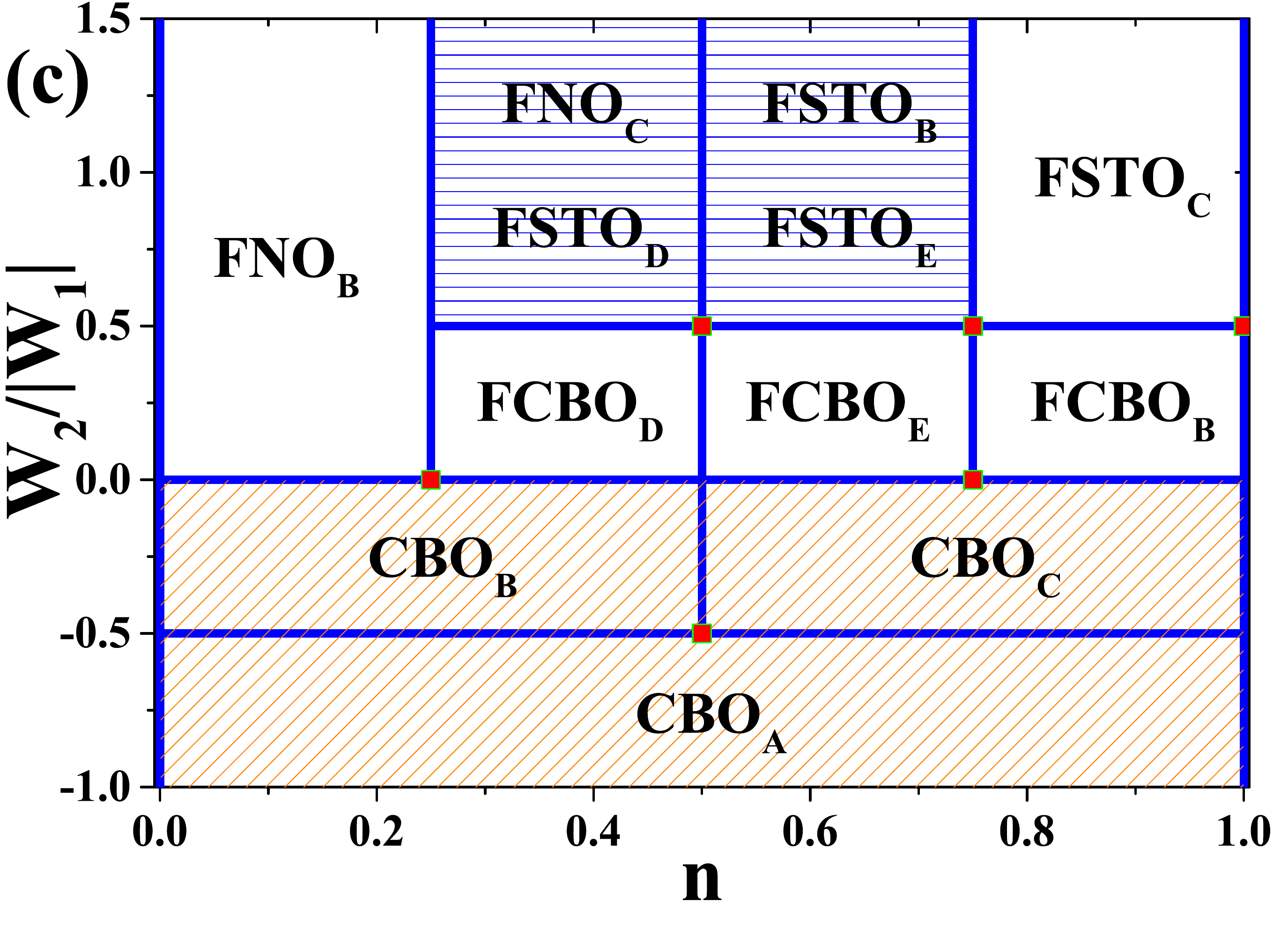}
	\includegraphics[width=\rozmiartrzy]{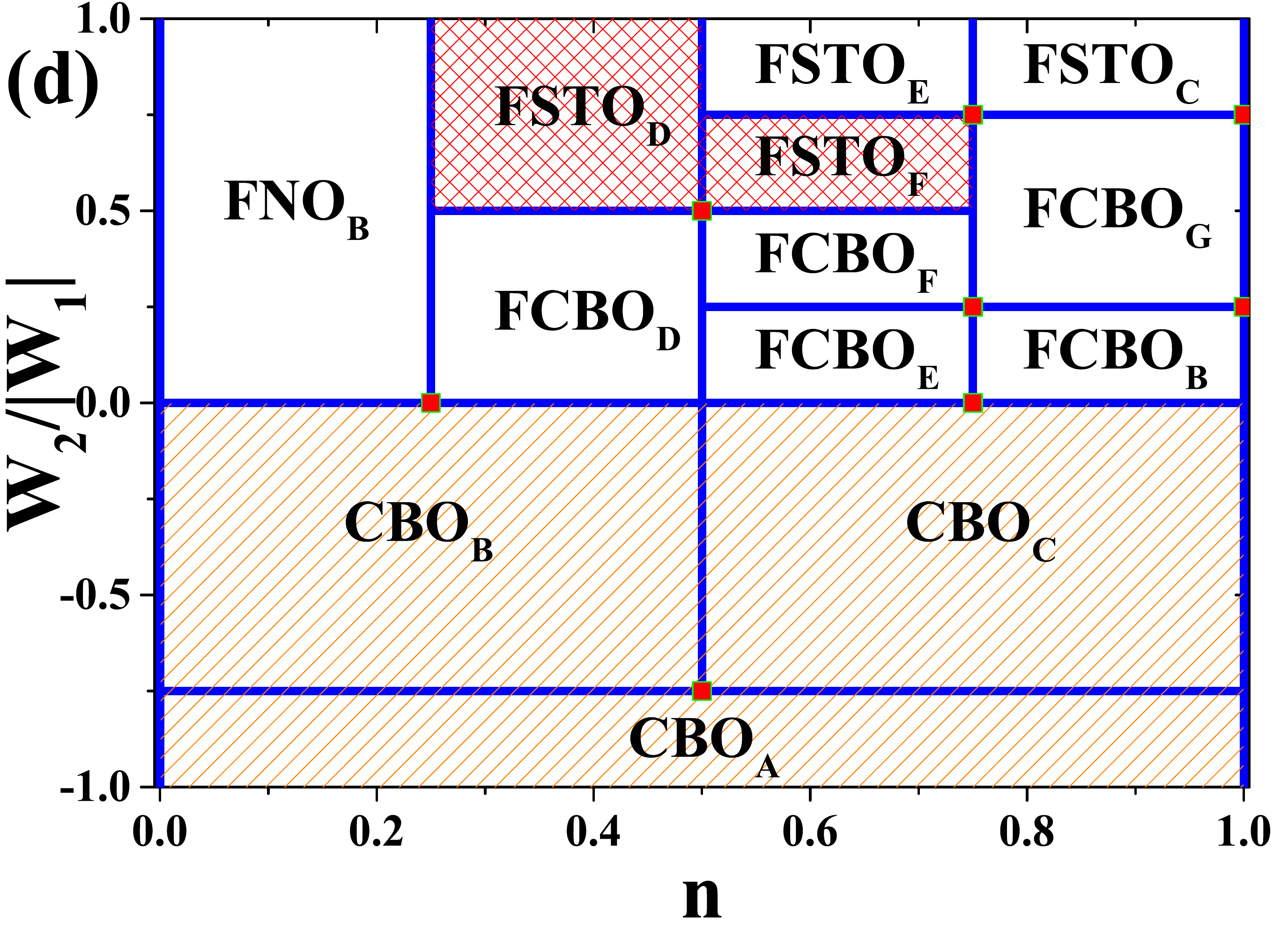}\\
	\includegraphics[width=\rozmiartrzy]{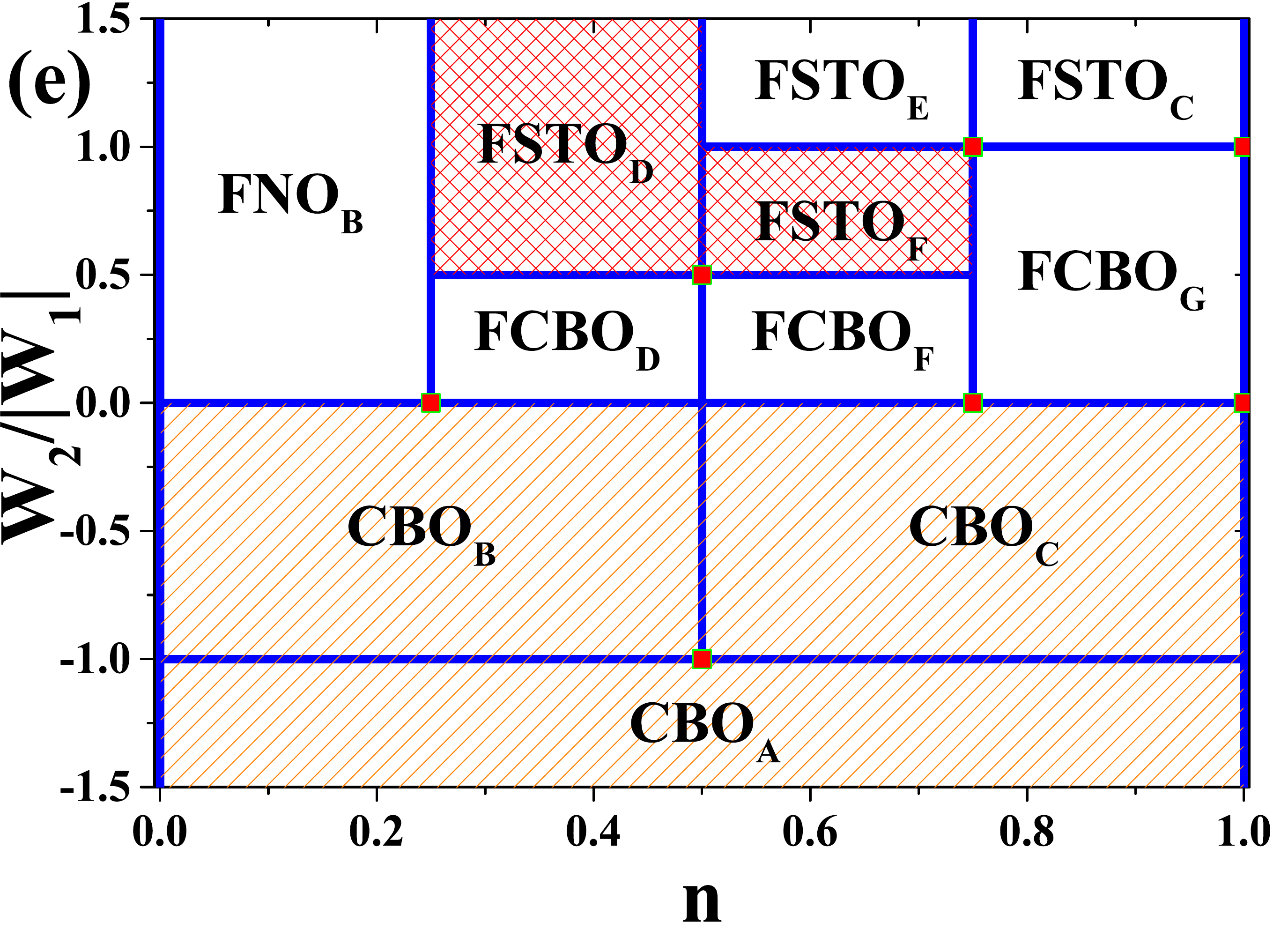}
	\includegraphics[width=\rozmiartrzy]{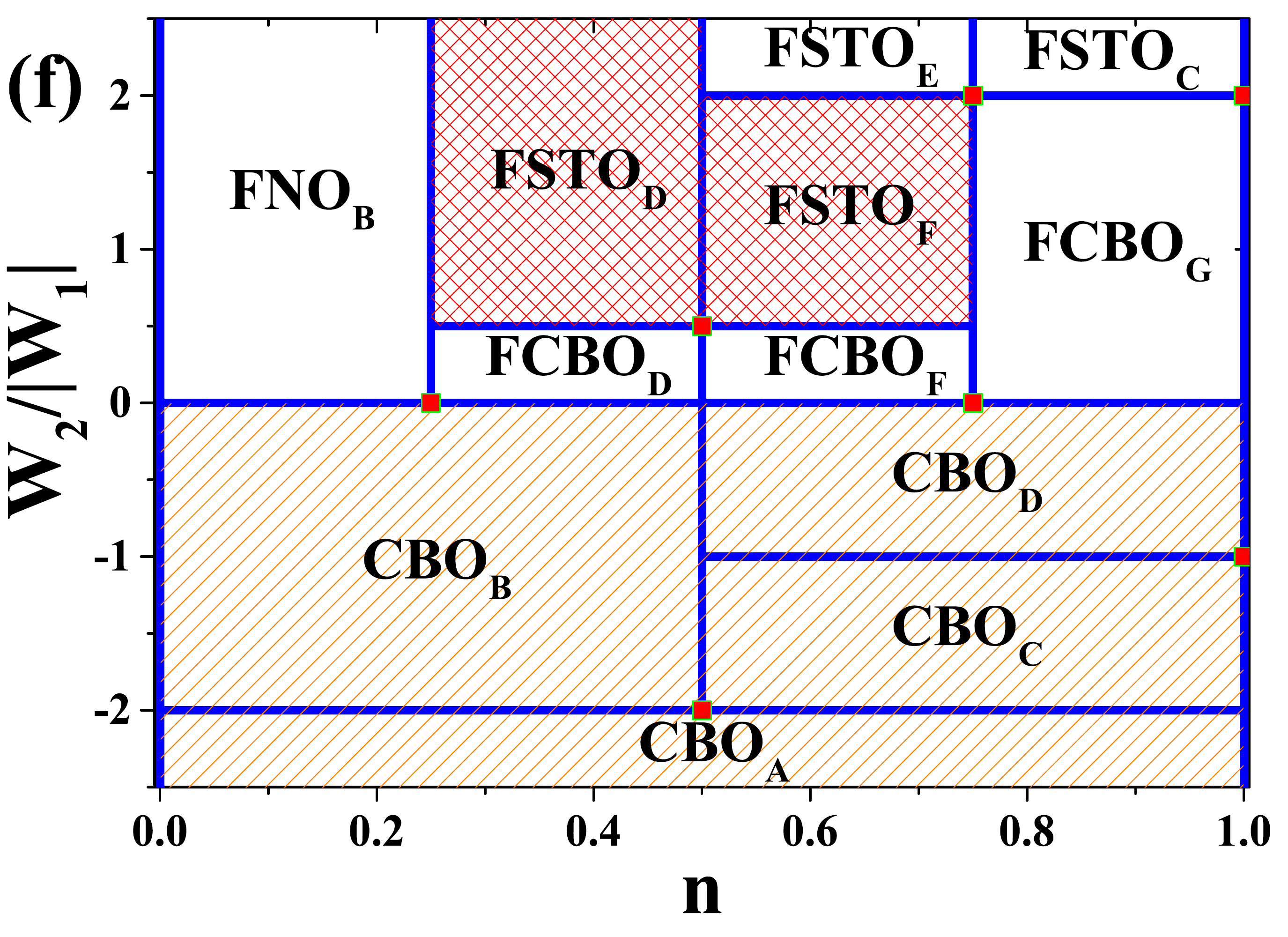}
	\caption{(Color online)
		Ground state phase diagrams as a function of electron concentration for $W_1>0$ and (a) $U/|W_1|\leq0.00$, (b) $U/|W_1|=0.25$, (c) $U/|W_1|=0.50$, (d) $U/|W_1|=0.75$, (e) $U/|W_1|=1.00$, (f) $U/|W_1|=2.00$.
		Each rectangular  region of the diagrams is labelled by the name of homogeneous phase with the lowest free energy (cf. Table~\ref{tab:concentartion}).
		In regions filled by slantwise pattern the corresponding PS state is stable at $T=0$ and at infinitesimally small $T>0$.
		In regions filled by grating pattern the homogeneous phase and PS state have equal free energies at $T=0$ whereas the PS state is stable at infinitesimally small $T>0$.
		In nonfilled regions  the homogeneous phase and PS state have equal free energies at $T=0$ whereas the homogeneous phase is stable at infinitesimally small $T>0$.
		At vertical boundaries only the homogeneous phases occur (cf. Table~\ref{tab:funkmi} and  Fig.~\ref{rys:GSchemical}).
		The rectangular points indicate discontinuous transitions at commensurate fillings.
	}
	\label{rys:GSconcnetration}
\end{figure*}

\subsection{Analysis for fixed electron concentration for repulsive $W_1$}

Here we consider the mutual relations between homogeneous phases and phase separated states for fixed total concentrations $n$.
To do this one needs to first establish which homogenous phases have the lowest energies and then compare them to the energies of phase separated states.
The first step is illustrated in Fig.~\ref{rys:GSconcnetration}, which presents the phase diagrams as a function of electron concentration.
Each rectangular  region of the diagrams is labeled by the abbreviation of homogeneous phase with the lowest free energy.
The meaning of each label is given in Table~\ref{tab:concentartion}.
The free energy per site of homogeneous phases at $T=0$ within the mean-field approximation can by obtained as $f_0 = UD_{occ}+E_W$, where $E_W$ is expressed by Eq. (\ref{eq:EW.tempzero}).
The double occupancy $D_{occ}$ and $n_\alpha$ are also collected in Table~\ref{tab:concentartion}.
On the vertical boundaries (for commensurate particle fillings $n=i/4$, $i=1,2,3,4$) the homogeneous phases occur.
These phases can be read from Table~\ref{tab:funkmi} and  Fig.~\ref{rys:GSchemical}).
For $n=1/2$ on the boundary between \FSTO$_\textrm{D}$ and \FSTO$_\textrm{F}$ regions the {\STOA} phase occurs  (Fig.~\ref{rys:GSconcnetration}(f)).
On the horizontal boundaries the phases from both neighboring regions have equal energies, and they coexist.

So far we labelled only homogenous phases with the lowest energies.
The comparison of them with the phase separated states shows that for $W_2<0$ the homogeneous phases are unstable (i.e. $\partial\mu / \partial n <0$) and thus only the PS states are present.
These regions in Fig.~\ref{rys:GSconcnetration} are marked by a slantwise pattern.
The obvious exclusion from this pattern occurs for vertical boundaries between two types of PS states where homogenous phases are stable below $W_2=0$.
We determined that at $T=0$ and $W_2 \geq 0$ homogenous phases and PS states have the same energy.
The corresponding PS states in each region of Fig.~\ref{rys:GSconcnetration} are given in the last column of Table~\ref{tab:concentartion}.
We would like to emphasize that the system cannot be simultaneously in a homogenous phase and a PS state.
Thus even though energies of PS states and homogeneous phases at $T=0$ are equal, the system must ``choose'' one of the solutions.
This type of degeneracy is removed by the finite temperature.
The free energies of the (macroscopic) PS states are calculated from the expression	$f_{PS}= mf_{+}(n_{+}) + (1 - m) f_{-}(n_{-})$, where $f_\pm(n_\pm)$ are free energies of separating homogeneous phases	and $m = (n - n_{-})/(n_{+}-n_{-})$ is a fraction of the system which is occupied by the phase with concentration $n_+$.

Additionally, we note that for fixed $W_2$ transitions between homogeneous phases (horizontal lines in Fig.~\ref{rys:GSconcnetration}) are associated with continuous change of sublattice concentrations, whereas for fixed $n$ (vertical boundaries) the sublattice concentrations change discontinuously
(at commensurate fillings only at points indicated by square symbols in Fig.~\ref{rys:GSconcnetration}).
All transitions between the phases exhibit discontinuous change of chemical potential.

Notice also that mean-field results presented in Fig.~\ref{rys:GSconcnetration} ($W_1>0$) are coincided with some exact results obtained for a one-dimensional chain at $T=0$ and arbitrary $n$.
In particular, the mean-field approximation used in the present work predicts properly the values for (free) energy, double occupancy correlation function, and nearest- and next-nearest two-point correlation functions \cite{ManciniPRE2008,ManciniEPJB2013}.
Ref.~\cite{ManciniPRE2008} predicts also that at $T=0$ and for $W_2=0$ ($W_1>0$) the long-range checker-board order exist in the following range of model parameters: (i) $0<U<W_1$ and $1/2<n<1$, (ii) $U>0$ and $n=1/2$, (iii) $U<W_1$ and $n=1$.
We found the similar behaviour for $W_2>0$ in the {\FCBO}$_{\textrm{B}}$ and {\FCBO}$_{\textrm{E}}$  regions (Fig.~\ref{rys:GSconcnetration}(b)--(d)), where the long-range checker-board order is expected for $d=1$ and $d=2$ at zero temperature.
For (a) $W_1>0$ and $W_2<0$ as well as (b) $W_1<0$ and any $W_2$ at $T=0$ and $d=1,2$  the macroscopic phase separation involving the checkerboard order would occur (for incommensurate fillings).

\subsubsection*{Influence of finite temperatures (fixed $n$)}

For $W_2<0$ the finite temperature does not change qualitative behavior of the system and the PS are still stable.
For $W_2\geq0$, infinitely small $T>0$ breaks the energy equality between homogeneous phases and PS states and only one of them is stable.
Namely, the PS states between {\FNO} and {\STO} phases emerge at the regions filled by grating pattern for $W_2>1/2$, whereas in empty regions in Fig.~\ref{rys:GSconcnetration} the homogeneous phases are favoured.
For $W_2=0$ only the homogeneous {\CBO} phases are present.
Remarkably, the regions of the PS states occurrence (slantwise and grating patterns) are separated by homogeneous phases (empty regions) at $T>0$.
The stability of the PS states at $T>0$ in these regions is a result of discontinuous transitions with changing $\bar{\mu}$, e.g., Refs.~\cite{KapciaJPCM2011,KapciaPhysA2016}.
Note that at $T>0$ the separating homogeneous phases are the phases with particle concentrations different than those at the ground state, and they can be determined by the so-called Maxwell's construction (cf., e.g., Ref.~\cite{KapciaJPCM2011}).

Due to the fact that at finite temperatures sublattice concentrations change continuously for fixed $W_2/|W_1|$ (vertical lines between nonfilled regions in Fig.~\ref{rys:GSconcnetration}) only boundaries between different types of phases still exist.
Boundaries between the same type of phases (e.g., \FCBO$_{\textrm{B}}$--\FCBO$_{\textrm{E}}$) are smeared out.
We also note that $T>0$ transforms all {\FNO} phases into {\FCBO} phase.
Therefore, ground state {\FNO}--{\FCBO} boundaries
are also smeared (vanish) at nonzero temperatures.
One should be aware of the type of phase occurring at commensurate filling (at vertical boundaries of diagrams in Fig.~\ref{rys:GSconcnetration}).
For example, at the ground state {\FCBO}$_{\textrm{B}}$--{\FCBO}$_{\textrm{E}}$ boundary (Fig.~\ref{rys:GSconcnetration}(c)) the {\FCBO} phase occur and thus this boundary vanishes at $T>0$.
The situation changes for   {\FCBO}$_{\textrm{D}}$--{\FCBO}$_{\textrm{E}}$ and {\FCBO}$_{\textrm{D}}$--{\FCBO}$_{\textrm{F}}$ boundaries (e.g. Fig.~\ref{rys:GSconcnetration}(d)), where for $n=1/2$ the {\CBOA} phase occurs at $T=0$.
As a result at small $T>0$ the {\FCBO}--{\CBO}--{\FCBO} sequence of transitions with changing $n$ is present.
For the ground state {\FNO}$_{\textrm{C}}$--{\FSTO}$_{\textrm{B}}$ boundary (Fig.~\ref{rys:GSconcnetration}(b)), the {\FNOB} phase is stable at $n=1/2$ and thus only the {\FCBO}--{\FSTO} transition occurs at finite $T$.
By using the scheme shown one can analyze each of the ground state boundary at $n=i/4$ ($i=1,2,3,4$).
One concludes that all ground state boundaries at $n=1/4$ and $n=3/4$ between homogeneous phases are smeared out at $T\neq0$, whereas all those for $n=0$ and $n=1$ remain.
For $n=1/2$ only the ground state {\FNO}$_{\textrm{A}}$--{\FCBO}$_{\textrm{A}}$  (Fig.~\ref{rys:GSconcnetration}(a)) and   {\FNO}$_{\textrm{C}}$--{\FCBO}$_{\textrm{C}}$ (Fig.~\ref{rys:GSconcnetration}(b)) vanishes at small finite $T$.

\begin{figure}[t!]
	\centering
	\includegraphics[width=\rozmiartrzy]{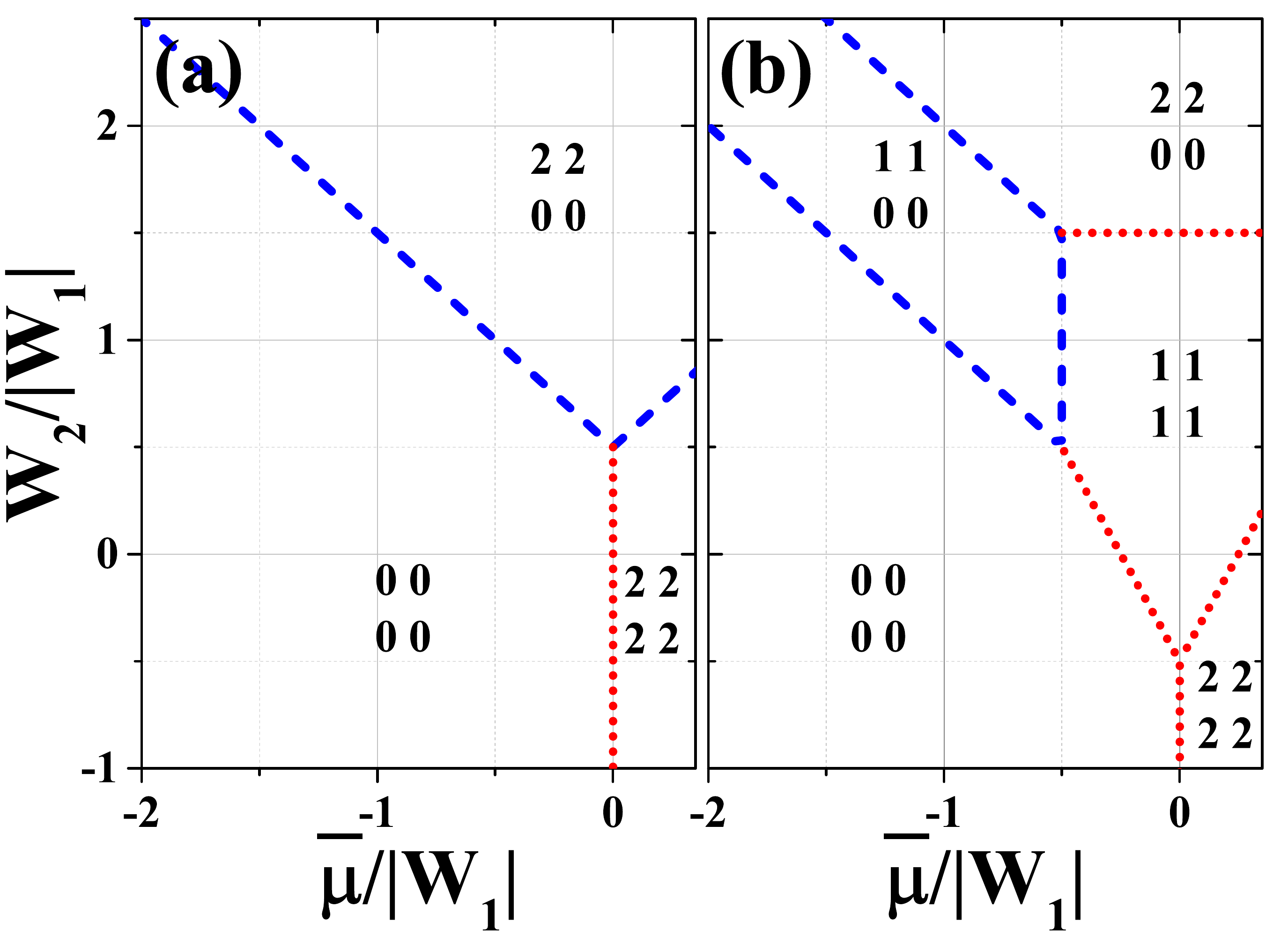}
	\caption{(Color online)
		Ground state phase diagrams as a function of chemical potential for $W_1<0$ and (a) $U/|W_1|\leq0$, (b) $U/|W_1|>0$ ($\bar{\mu} = \mu- \tfrac{1}{2}U-W_1-W_2$).
		Each region is labelled by electron concentrations in each sublattice: 	$\left( \begin{smallmatrix} n_A & n_B \\ n_D & n_C \end{smallmatrix} \right)$ (cf. Table~\ref{tab:funkmi}).
		Dashed lines denote infinite degeneration but not macroscopic, whereas dotted lines indicate finite degeneration (at the boundaries for $d=2$).
	}
	\label{rys:W1m0mi}
\end{figure}

\begin{figure}[t!]
	\centering
	\includegraphics[width=\rozmiartrzy]{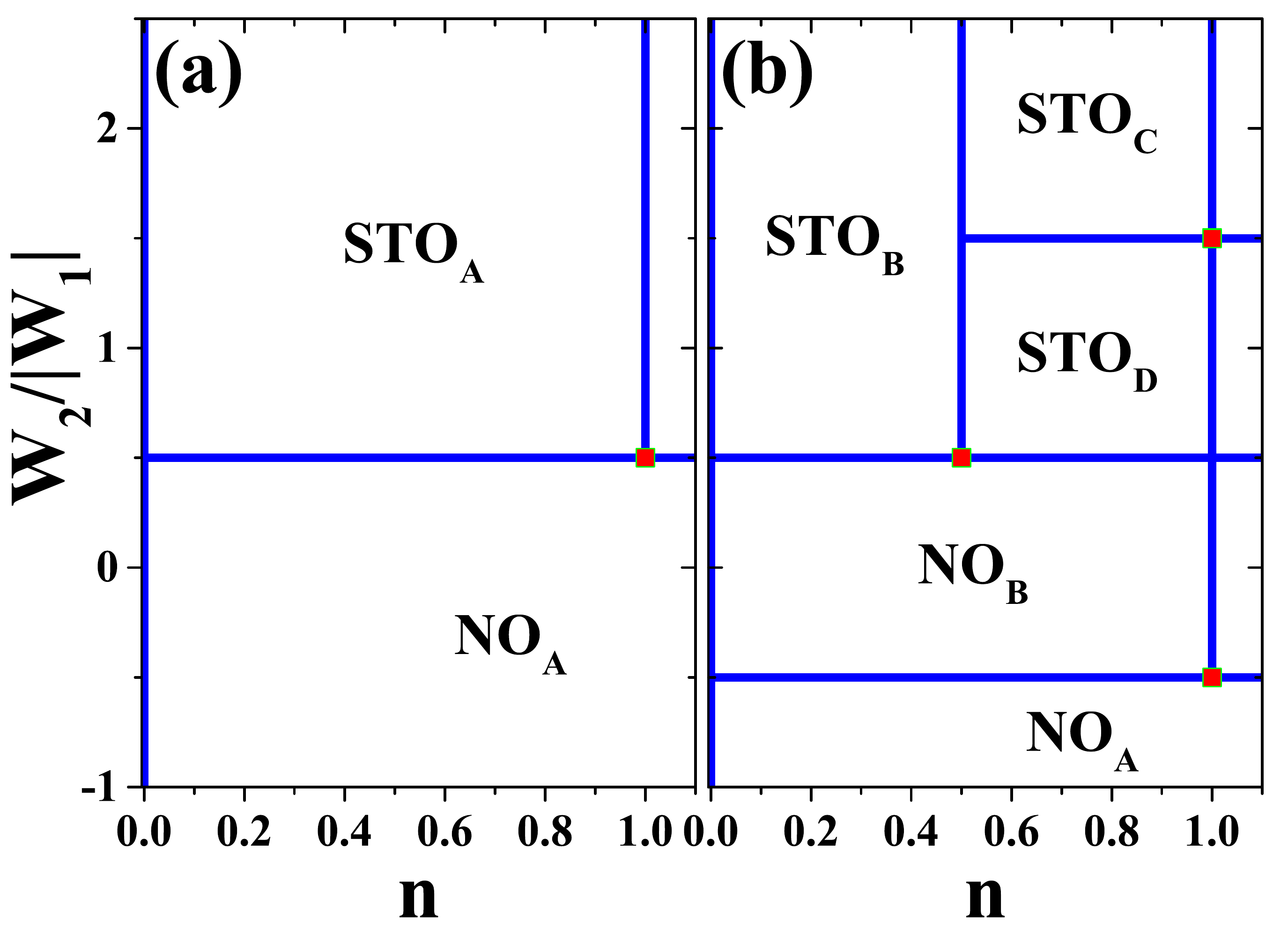}
	\caption{(Color online)
		Ground state phase diagrams as a function of electron concentration for $W_1<0$ and  (a) $U/|W_1|\leq0$, (b) $U/|W_1|>0$.
		Each rectangular  region of the diagrams is labelled by the name of the homogeneous phase with the lowest free energy (cf. Table~\ref{tab:concentartion}).
		In all regions the corresponding PS states are stable at $T=0$ and at infinitesimally small $T>0$ (excluding vertical boundaries).
		The rectangular points indicate discontinuous transitions at commensurate fillings (cf. Fig.~\ref{rys:GSchemical}).
	}
	\label{rys:W1m0n}
\end{figure}

Finally, let us comment on transitions with changing $W_2/|W_1|$ for fixed $n$.
All horizontal ground state boundaries, for $W_2\neq0$ remain discontinuous at $T>0$.
At the specific case of $W_2=0$ the  {\CBO}--{\FCBO} and {\CBO}--{\FNO} boundaries are second order at nonzero temperatures (cf. also Ref. \cite{KapciaJSNM2016}).
Moreover, the ground state {\CBO}$_{\textrm{A}}$--{\CBO}$_{\textrm{B}}$ and {\CBO}$_{\textrm{A}}$--{\CBO}$_{\textrm{C}}$ boundaries vanish at any finite $T$ for $W_2=0$ (cf. Refs. \cite{MicnasPRB1984,KapciaJPCM2011,KapciaPhysA2016}).

\subsection{The case of attractive $W_1$}

To provide the full picture of the system ground state we present results obtained for the  attractive nearest-neighbour interaction ($W_1<0$).

This case is less complex than a case of $W_1$ as the qualitative behaviour of the system is not dependent on a magnitude of the on-site interaction (only the sign of $U$ is relevant).
For attractive $W_1$  only possible phases are either nonordered or stripe-ordered (\STO) phases.
In the case of $U<0$ only phases with empty or double-occupied sites are realized (Fig.~\ref{rys:W1m0mi}(a)), whereas for $U>0$ the phases with single-occupied sites emerge (Fig.~\ref{rys:W1m0mi}(b)).
All phase boundaries are discontinuous.
The above behaviour of the system is preserved also at small, but finite temperatures.

Similarly to the previous analyses, in Fig.~\ref{rys:W1m0n} we present the phase diagrams as a function of concentration $n$.
In this case the corresponding PS states (cf. Table~\ref{tab:concentartion}) possess lower energies than homogeneous phases in the full range of the model parameters (excluding vertical boundaries in Fig.~\ref{rys:W1m0n}).
The attractive $W_2$ favours the PS states between nonordered phases, whereas repulsive $W_2$ gives rise to PS states involving various {\STO} phases.
Homogenous phases on the vertical boundaries ($n=i/2$, $i=1,2$) can be read  from  Fig.~\ref{rys:W1m0mi} and Table~\ref{tab:funkmi}.
On the horizontal boundaries the PS states from both neighboring regions have equal energies and they coexist.
Moreover, at infinitesimally $T>0$ the diagrams do not change and the PS states  still occur.

\section{Low dimensional systems}
\label{sec:lowdim}

So fa we have discussed the mean-field solutions of the model, which is an exact theory in high dimensions (formally $d\to+\infty$, $z_n\to +\infty$) \cite{MullerHartmannZPB1989,PearceCMP1975,PearceCMP1978}.
In this section we discuss the qualitative behavior of lower dimension systems, which fulfill the four-sublattice restriction.
In particular, we consider the 1D chain, 2D square (SQ) lattice, and 3D base-centered cubic (BCC) lattice.
For the 3D  one we choose the BCC lattice because (unlike simple cubic or face-centered cubic lattices) it can be divided into four equivalent sublattices.

A set of four sublattice concentrations ($n_An_Bn_Cn_D$) defines an elementary block.
These blocks can be grouped in types.
All blocks of a given type can be obtained by a cyclic change of sublattice indices,
e.g., the {\FCBO} type consists of four different elementary blocks, namely [(2010), (0201), (1020), (0102)].
Any elementary block of a ground state configuration must be among those that yield the minimal energy when extended periodically \cite{Pirogov1975,Pirogov1976,FrohlichCMP1978,FrohlichJSP1980,JedrzejewskiPhysA1994}.
In other words, in a construction of a $T=0$ diagram as a function of $\bar{\mu}$ only elementary blocks with the lowest energy and periodic configurations obtained from them play a role.
The  grand canonical potentials of such states at $T=0$ are equal to the potentials of mean-field phases determined  from Eqs. (\ref{eq:grandpotential.tempzero})--(\ref{eq:Emu.tempzero}) and collected in Table~\ref{tab:funkmi}.
Thus the structure of the ground state diagrams presented in Figs.~\ref{rys:GSchemical} and~\ref{rys:W1m0mi} remains unchanged in lower dimensions, but the properties of the system at the boundaries and inside given regions can change.

Let us look closely at conditions at which at least two phases (elementary blocks) have equal energies, i.e., at the boundaries between regions on the phase diagrams as a function of chemical potential.
In dimensions $d\geq3$ any two elementary blocks of different phases can not be mixed with each other.
In lower dimensions $d<3$ this mixing become possible and additional degeneracy appears.
Generally, we can classify three types of phase boundaries with respect to elementary blocks mixing:

\begin{figure}[t!]
	\centering
	\includegraphics[width=0.25\textwidth]{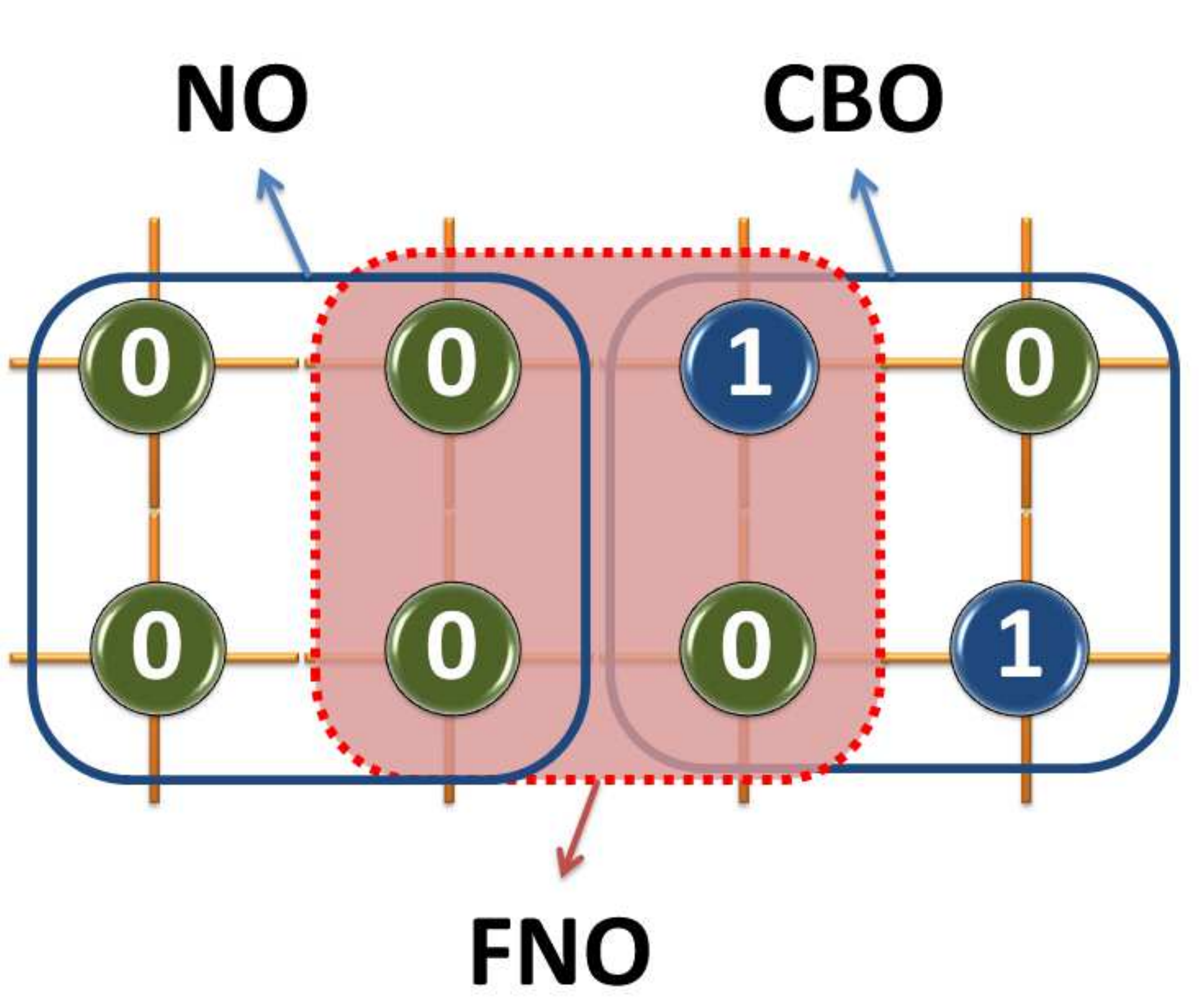}
	\caption{(Color online)
		Schematic illustration  of \emph{nonmixing} block alignment in $d=2$.
		Note that the {\FNO} block at the interface between the {\NO} and {\CBO} phases emerge (shadow, dotted line).
		For $W_2/|W_1|<0$ the {\FNO} block does not yield the minimal energy.
	}
	\label{rys:schem1}
\end{figure}
\begin{figure}[t!]
	\centering
	\includegraphics[width=0.23\textwidth]{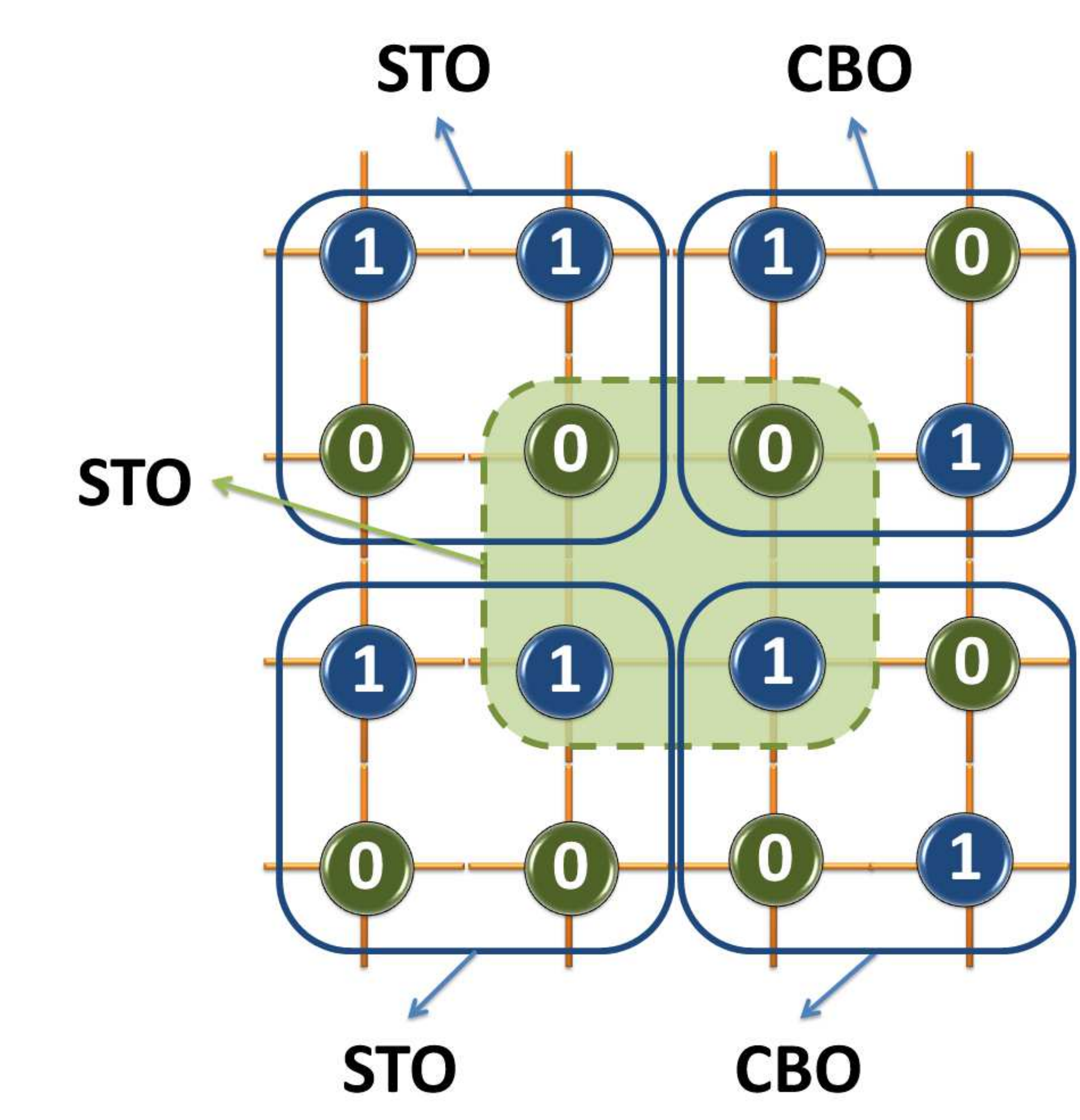}
	\includegraphics[width=0.23\textwidth]{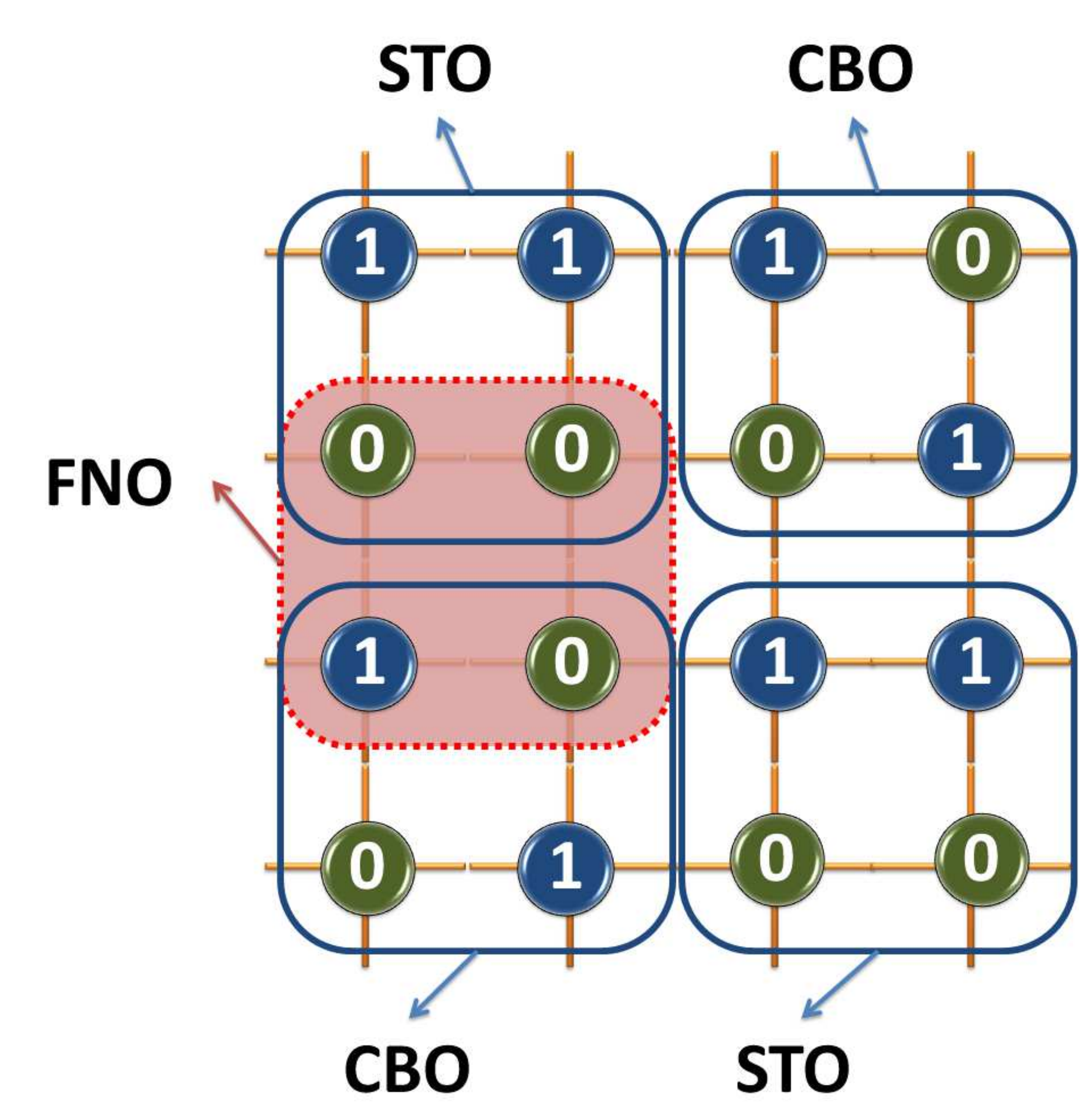}
	\caption{(Color online)
		Schematic illustration  of \emph{partial-mixing} block alignments in $d=2$ at the {\STO}--{\CBO} interface.	
		In the left panel an allowed block orientation is shown (at the interfaces the {\STO} or {\CBO} blocks emerge).		
		In the right panel a forbidden configuration is present, where in addition the {\FNO} blocks emerge at the interface.
	}
	\label{rys:schem2}
\end{figure}

\begin{figure}[t!]
	\centering
	\includegraphics[width=0.25\textwidth]{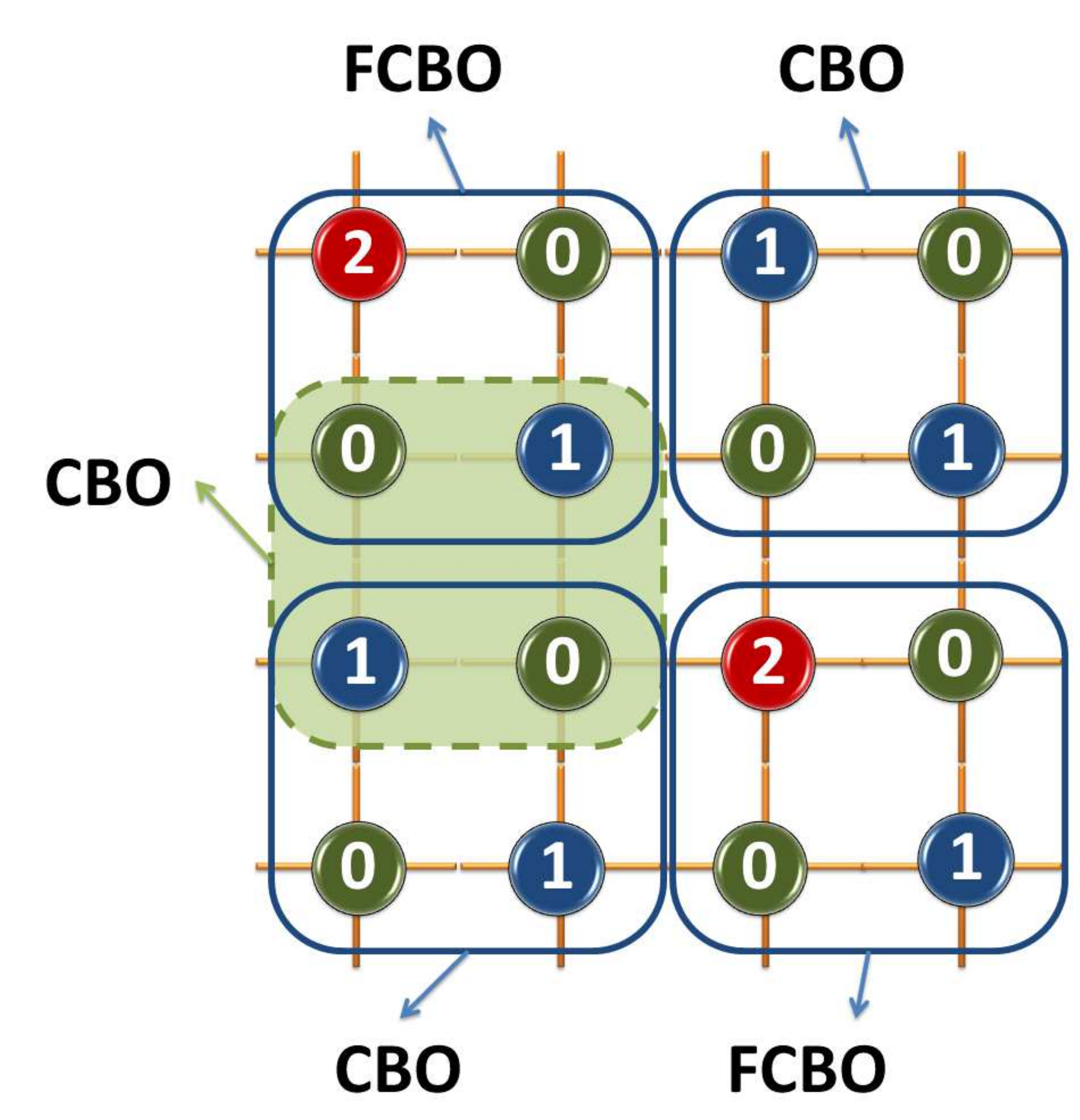}
	\caption{(Color online)
		Schematic illustration  of \emph{full-mixing} block alignment in $d=2$.
		All building blocks at the interface between the {\CBO} and {\FCBO} phases are blocks of one of these mixing phases.
		An exemplary block at the interface is denoted by the shadow (dashed line).
	}
	\label{rys:schem3}
\end{figure}

\begin{itemize}
	\item [(i)] \emph{Nonmixing case}.
	In this case on the boundaries between two phases the state of the system can be constructed by periodical repetition of only one type of elementary block.
	In other words elementary block of one type cannot be aligned next to a block of a different type.
	As an example of such a situation let us consider the {\NOA}--{\CBOA} boundary.
	The elementary blocks of these phases are respectively
	$\begin{smallmatrix} 0 & 0\\ 0 & 0 \end{smallmatrix}$
	and
	$\begin{smallmatrix} 1 & 0\\ 0 & 1 \end{smallmatrix}$.
	In Fig.~\ref{rys:schem1} we show that if such blocks would be aligned next to each other  the region
	with block
	$\begin{smallmatrix} 0 & 1\\ 0 & 0 \end{smallmatrix}$ would always be created
	that does not yield the minimal energy.
	In such a case even though both phases possess equal energies the ground state cannot be built of the composition of them, but it must ``choose'' one of the solutions.
	Then a coexistence of the phases can be realized only on a macroscopic level and only if the contribution of the domain wall energy vanishes in the thermodynamical limit (the so-called macroscopic phase separation).
	Such types of boundaries at $d=2$ are denoted by dotted lines in Figs.~\ref{rys:GSchemical} and~\ref{rys:W1m0mi}.
	At the boundaries the degeneracy is finite (modulo spin).
	\item[(ii)]\emph{Partial mixing.}
	In Figs.~\ref{rys:GSchemical} and~\ref{rys:W1m0mi} dashed-lines denote boundaries at which elementary blocks of  different types can be aligned next to each other, but not arbitrarily (at $d=2$).
	Some restrictions on a block's configuration remains.
	An example of this type of the boundary occurs between the {\CBOA}  and  {\STOA} phases (see Fig.~\ref{rys:schem2}).
	We notice that if
	$\begin{smallmatrix} 1 & 0\\ 0 & 1 \end{smallmatrix}$
	and
	$\begin{smallmatrix} 1 & 1\\ 0 & 0 \end{smallmatrix}$
	blocks are aligned next to each other (in a given direction) we cannot find any region built of elementary blocks which does not belong to one of the phases with minimal energy ({\CBOA} or {\STOA}).
	Specifically, it means that every column (or row) of blocks has to be purely built of one type of elementary blocks, whereas blocks can be freely aligned in rows (or columns).
	We note that opposed to nonmixing regime here we get microscopical mixing of each phase, but some macroscopic ordering remains as rows (or columns) are build of one type of elementary blocks.
	At such boundaries degeneracy  $\Gamma$ of the system is infinite but not macroscopic (modulo spin), i.e., it increases with size of the system (i.e., with $L$) lower than $bA^L$ (where $b$ and $A$ are some fixed numbers; $0<A<3$) (entropy per site in the thermodynamic limit $s=\lim_{L\rightarrow+\infty}\tfrac{1}{L}\ln \Gamma$ is zero).
	Such boundaries at $d=2$ are denoted by dashed lines in Figs.~\ref{rys:GSchemical} and~\ref{rys:W1m0mi}.
	\item[(iii)] \emph{Full mixing.}
	Solid lines in Fig.~\ref{rys:GSchemical} denote boundaries at which elementary blocks of neighboring regions can be aligned freely without any restrictions for $d=2$ (macroscopic degeneration).
	As an example one can consider the {\FCBO-\CBOA} boundary; see Fig.~\ref{rys:schem3}.
	In this case the system is microscopically mixed and therefore no macroscopic orderings are present.
    This type of boundaries at $d=2$ is denoted by solid lines in Figs.~\ref{rys:GSchemical}.
    At such boundaries degeneracy $\Gamma$ is infinite and macroscopic (modulo spin), i.e., it increases faster than  $bA^L$ (entropy per site in the thermodynamic limit $s=\lim_{L\rightarrow+\infty}\tfrac{1}{L}\ln \Gamma$ is finite).
\end{itemize}
For $d=1$ the boundaries with infinite degeneration in Figs.~\ref{rys:GSchemical} and~\ref{rys:W1m0mi} (dashed and solid lines) are macroscopically degenerated.
In such a dimensionality of the system there is no \emph{partial-mixing} at the boundaries.

Away from boundaries, in regions filled by the slantwise pattern in Fig.~\ref{rys:GSchemical},
two configurations of elementary blocks of the same phase can be mixed with each other, but arbitrarily.
This situation is similar to a case of the \emph{partial mixing} at phase boundaries.
In low dimensions ($d=1,2$) mixing of the same type of elementary blocks but in different configurations (e.g.
$\begin{smallmatrix} 2 & 0\\ 1 & 0 \end{smallmatrix}$ and $\begin{smallmatrix} 1 & 0\\ 2 & 0 \end{smallmatrix}$) leads to increased disorder in contrary to $d=3$ case, where one configuration is present in whole system.
Analogously to \emph{partial mixing}, in the discussed cases every column (row) of blocks have to be purely built of elementary blocks in the same configurations, but blocks can vary from row to row (column to column), cf. Fig.~\ref{rys:schem2}.
The ground state for $d=2$ marked by the slantwise pattern in Fig.~\ref{rys:GSchemical} has infinite degeneration, but it is not macroscopic (cf. Table~\ref{tab:funkmi}).
For $d=1$ the degeneracy in these regions is macroscopic.

To reach the situation, where in these regions a long-range charge-order would be present, an arbitrary weak interaction  between third-nearest neighbors ($W_3\neq0$) is sufficient.
$W_3<0$ gives long-range order of the four-sublattice type  for SC and BCC lattices (for $W_3>0$ $4\times 2$ (eight-sublattice) orderings need to be considered) \cite{LandauPRB1985,LiuPRE2016}.
In a case of  1D chain the long-range order appears for $W_3>0$, whereas $W_3<0$ stabilizes eight-sublattice orderings.

The phases with single-occupied sites are infinitely (macroscopically) degenerate with respect to spin degrees of freedom and thus the model considered does not exhibit any magnetic order in any dimension.

Similarly as the Ising model with short-range interactions,  model (\ref{eq:hamUW})  considered on one-dimensional lattice does not exhibit long-range order at any $T>0$ for any model parameters \cite{ManciniPRE2008,ManciniEPJB2013}.
For the two-dimensional lattice the order above the ground state {\CBO}, {\STO} and {\FCBO}, regions (i.e., nonfilled regions in Fig.~\ref{rys:GSchemical}) would occur for fixed $\bar{\mu}$ \cite{JędrzejewskZPhysB1985,JedrzejewskiPhysA1994,BorgsJPA1996,BinderPRB1980,AnjosPRE2007,BobakPRB2015}.
At incommensurate fillings one should expect the order at $T>0$ also in the {\FCBO}$_{\textrm{B}}$ and {\FCBO}$_{\textrm{E}}$  regions as well as in regions where the PS states are stable at $T=0$.
For $d\geq3$ the order should be present at small $T>0$ for all model parameters (excluding those for which the {\NO} phase occurs at $T=0$).

\section{Final remarks}

We considered the zero-bandwidth extended Hubbard model taking into account longer-range (NN and NNN) interactions.
It was found that such correlations give rise to four-sublattice solutions for $W_2>0$ at the mean-field level (with exact treatment of the on-site term).
We indicated that these new phases emerge at a given magnitude of the on-site interactions $U$.
It was shown that the {\FNO} phase is present for arbitrary values of $U$, and the {\FSTO} phase occurs only for repulsive $U$, while the {\FCBO} phase is limited by the $0<U<W_1$ condition.
For fixed electron concentration the system is highly degenerated, but arbitrary small temperature ($T>0$) removes this degeneracy.
We also show that not all phase transitions occurring at $T=0$ remain first-order at finite temperatures.
They are second-order ones at $T>0$.

We discussed the influence of the lattice dimensionality $d$ on the degeneracy of the ground state (for fixed $\mu$).
For the $d=3$ lattice the results are in an agreement with the mean-field findings and finite degeneracy is present for arbitrary model parameters.
In a case of lower dimensionalities ($d=1,2$) the  appearance of the partial and full mixing of the elementary blocks gives rise to infinite degeneracy.
We showed also that four-sublattice long-range order in low dimensions is suppressed (in the {\FNO} and {\FSTO} phases), but in the {\FCBO} phase the long-range order remains.

Finally, let us note that in this work we analyzed the system neglecting the influence of the electron hopping term.
It is known that finite hopping induces additional magnetic orderings \cite{NagaokaPRB1966,Chao1977,Chao1978,PencPRB1994,Giovannetti2015}.
The other aspect related to quantum fluctuations introduced by the hopping is a metal-insulator transition.
Basing on the results obtained for two sublattice assumption (for $W_2=0$) \cite{AmaricciPRB2010,HuangPRB2014,Giovannetti2015,KapciaPRB2017} one can expect that {\CBO} and {\STO} phases for $W_2\neq0$ should survive in a presence of such fluctuations.
It is an open question whether other orderings (i.e.,  {\FNO}, {\FCBO}, and {\FSTO}) will also remain (in insulating or metallic states).
Although our system is simplified for this point, the results obtained here are exact solutions.
Therefore, they can be used to inspect the validity of approximations used to the more general models including single-electron hopping.

\begin{acknowledgments}
    The authors thank Anna Ciechan for very fruitful discussions.
	The authors acknowledge the support from  the National Science Centre (NCN, Poland) under grants no. UMO-2014/15/B/ST3/03898 (K.J.K), UMO-2016/21/D/ST3/03385 (K.J.K. and J.B.), UMO-2017/24/C/ST3/00276 (K.J.K.), and UMO-2016/20/S/ST3/00274 (A.P.).	
\end{acknowledgments}

\appendix

\section{The equivalent models}
\label{app:equivmodels}

One can show (cf. e.g. Refs.~\cite{JedrzejewskiPhysA1994,PawlowskiEPJB2006}) that model (\ref{eq:hamUW}) is equivalent with the classical Blume-Capel model with spin $S=1$~\cite{BlumePR1966,CapelPhys1966,BlumePRA1971,BadehdahEPJB1998} in the external magnetic field, which has the following form
\begin{equation}\label{eq:hamBC}
\hat{H}_{BC}=\Delta\sum_i \left(\tilde{S}^z_i\right)^2 + \frac{1}{2}\sum_{i,j}J_{ij}\tilde{S}^z_i \tilde{S}^z_j - H \sum_i\tilde{S}^z_i + C,
\end{equation}
where $\Delta = \tfrac{1}{2}U+ k_BT \ln(2)$ is temperature-dependent single-ion anisotropy, $J_{ij}=W_{ij}$, $H \equiv \bar{\mu} = \mu - U/2 - \sum_n z_n W_n$, $C = L \left(k_BT \ln(2) + \mu \right)$.
For $U\rightarrow-\infty$ models (\ref{eq:hamUW}) and (\ref{eq:hamBC}) are reduced to the standard $S=1/2$ Ising model ($S_i^z=\pm1$)~\cite{YamadaPTP1967,KatsuraJPC1974,KincaidPysRep1975,BinderPRB1980,AnjosPRE2007,BobakPRB2015}.
At $T=0$ for $U=0$ models (\ref{eq:hamUW}) and (\ref{eq:hamBC}) are reduced to the $S=1$ Ising model ($S_i^z=-1,0,1$).
In Eq.~(\ref{eq:hamBC}) we do not restrict the range of intersite interactions.


\bibliography{biblio_CO}


\end{document}